\documentclass[12pt]{article}
 \usepackage{epsfig}
 \usepackage{pstricks}
 \usepackage{pst-coil}
                                       
\newcommand{\dd}{\mbox{d}}

\def\a{\alpha}

\def\c{\chi}
\def\d{\delta}
\def\e{\epsilon}
\def\f{\phi}
\def\g{\gamma}

\def\j{\psi}

\def\l{\lambda}
\def\m{\mu}
\def\n{\nu}

\def\p{\pi}
\def\q{\theta}
\def\r{\rho}
\def\s{\sigma}
\def\t{\tau}

\def\z{\zeta}

\def\G{\Gamma}

\def\ve{\varepsilon}
\def\vf{\varphi}

\def\cl{{\cal L}}

\def\co{{\cal O}}

\def\bo{{\raise.15ex\hbox{\large$\Box$}}}               
\def\pr{\prod}                                          
\def\face{{\raise.2ex\hbox{$\displaystyle \bigodot$}\mskip-2.2mu \llap {$\ddot
        \smile$}}}                                      
\def\dg{\dagger}                                     

\def\Slash#1{\rlap{\hbox{$\mskip 3 mu /$}}#1}      
\def\VEV#1{\left\langle #1\right\rangle}        
\def\abs#1{\left| #1\right|}                    
\def\leftrightarrowfill{$\mathsurround=0pt \mathord\leftarrow \mkern-6mu
        \cleaders\hbox{$\mkern-2mu \mathord- \mkern-2mu$}\hfill
        \mkern-6mu \mathord\rightarrow$}       
\def\dvec#1{\vbox{\ialign{##\crcr
        \leftrightarrowfill\crcr\noalign{\kern-1pt\nointerlineskip}
        $\hfil\displaystyle{#1}\hfil$\crcr}}}           

\def\beq{\begin{equation}}
\def\eeq{\end{equation}}

\def\beqx{\begin{displaymath}}
\def\eeqx{\end{displaymath}}

\def\beqa{\begin{eqnarray}}
\def\eeqa{\end{eqnarray}}
\def\NO{\nonumber}

\def\pl#1#2#3{Phys.~Lett.~{\bf B {#1}} (19{#2}) #3}
\def\np#1#2#3{Nucl.~Phys.~{\bf B {#1}} (19{#2}) #3}

\def\pr#1#2#3{Phys.~Rev.~{\bf D {#1}} (19{#2}) #3}

\begin{document}
\date{}

\title{ 
{\normalsize     
\hfill \parbox{50mm}{DESY 96-052\\hep-ph/9604229}}\\[25mm]
Baryogenesis and   \\
Lepton number violation   \\[8mm]}
\author{
Michael Pl\"umacher\thanks{e-mail: pluemi@x4u2.desy.de}\\
Deutsches Elektronen-Synchrotron DESY,      \\
Notkestr.\,85, D-22603 Hamburg, Germany}
\maketitle

\begin{abstract}
  \noindent
  The cosmological baryon asymmetry can be explained by the
  nonperturbative electroweak reprocessing of a lepton asymmetry
  generated in the out-of-equilibrium decay of heavy right-handed
  \mbox{Majorana} neutrinos. We analyze this mechanism in detail in
  the framework of a SO$(10)$-subgroup. We take three right-handed
  neutrinos into account and discuss physical neutrino mass matrices.
\end{abstract}

\newpage  

\section{Introduction}
    One of the most striking features of the observable universe is
    the baryon asymmetry, which is usually expressed as the ratio of
    the baryon density $n_B$ to the entropy density $s$. From
    measurements of the abundances of the light elements one 
    finds\footnote{For a review and references, see \cite{kt1}}:
    \beq
    Y_B={n_B\over s}=(0.6-1)\times10^{-10}\;.
    \eeq
    With the appearance of grand unified theories (GUTs) it became
    possible to explain this asymmetry by the baryon number $(B)$
    violating decays of Higgs or gauge bosons at the GUT scale.
    However, these models of baryogenesis are not easily reconciled
    with inflation. Indeed, a baryon asymmetry present before
    inflation would be diluted by a huge factor, while the reheating
    temperature after the inflationary phase is in general too low for
    these baryogenesis mechanisms to work.

    Then, it was realized that anomalous baryon number violating
    processes are unsuppressed at high temperatures \cite{sphal2}.
    These so called sphaleron transitions violate $(B+L)$ and conserve
    $(B-L)$, where $L$ is the lepton number. As sphalerons are in
    thermal equilibrium for temperatures between $\sim\!\!10^{12}\,$GeV
    and $\sim\!\!10^2\,$GeV, they will strongly modify any primordial
    $(B+L)$ asymmetry. The connection between the baryon asymmetry and
    a primordial $(B-L)$ asymmetry is given by \cite{sphal}
    \beq
     Y_B=\left({8N_f+4N_H\over22N_f+13N_H}\right)Y_{B-L}\;,
    \eeq
    where $N_f$ is the number of fermion families and $N_H$ is the
    number of Higgs doublets.

    The needed primordial $(B-L)$ asymmetry can be realized as a
    lepton asymmetry generated by the out-of-equilibrium decay of
    heavy right-handed Majorana neutrinos, as suggested by Fukugita
    and Yanagida \cite{fy2}.  $L$ is violated by Majorana masses,
    while the necessary $CP$ violation comes about through phases in
    the Dirac mass matrix of the neutrinos. In a detailed
    quantitative analysis Luty showed that the scenario works for a
    wide range of parameters \cite{luty}.

    In order to generate a lepton asymmetry of the correct order of
    magnitude, the right-handed neutrinos have to be numerous before
    decaying. This is only possible if they are in thermal equilibrium
    at high temperatures. In the original model \cite{fy2} the
    right-handed neutrinos were only interacting through Yukawa
    couplings, which are far too weak to bring the neutrinos into
    equilibrium at high temperatures.  Hence one had to assume an
    equilibrium distribution for the neutrinos as initial condition in
    previous analyses.

    An appealing way to solve this problem is to study this
    baryogenesis mechanism in the framework of an extended gauge
    symmetry, since right-handed neutrinos appear naturally in
    unified theories based on the gauge groups SO$(10)$ or E$_6$. As
    we shall see, the gauge interactions in which the right-handed
    neutrinos take part, are strong enough to bring them into thermal
    equilibrium at high temperatures. Of course, the neutrinos have to
    be out of equilibrium when decaying, i.e.\ the reaction rates for
    the gauge interactions have to fall fast enough, so that they
    cannot significantly reduce the number density of the neutrinos 
    before they decay.

    In this paper we investigate this mechanism in the framework of an
    SO$(10)$ subgroup. After a short discussion of the relevant
    Boltzmann equations in the next section, we present our model and
    calculate the needed reaction rates in section \ref{model}. In
    section \ref{solutions} we solve the Boltzmann equations, first
    for one and then several heavy neutrino families. We explicitly
    show that the lepton asymmetry is mainly determined by the
    lightest of the right-handed neutrinos. Finally we will look at
    physical mass matrices for the neutrinos coming from an additional
    abelian gauged family symmetry. These mass matrices give a lepton
    asymmetry in the right order of magnitude and predict light
    neutrino masses and mixings of the magnitude needed to explain the
    solar neutrino deficit and a $\t$-neutrino mass of a few eV which
    is needed in the cold-plus-hot dark matter models.
\section{Boltzmann Equations \label{kinetic}}
    In a quantitative analysis of baryogenesis one can assume
    Maxwell-Boltzmann statistics \cite{kw}, so that the equilibrium
    phase space density of a particle $\j$ with mass $m_{\j}$ is given
    by
    \beq
     f^{eq}_{\j}\left(E_{\j},T\right)=\mbox{e}^{-E_{\j}/T}\;.
    \eeq
    The particle density is
    \beq
     n_{\j}(T)={g_{\j}\over(2\p)^3}\int\dd^3p_{\j}\,f_{\j}\;,
    \eeq
    where $g_{\j}$ is the number of internal degrees of freedom. The
    number of particles $Y_{\j}$ in a comoving volume element is given
    by the ratio of $n_{\j}$ and the entropy density $s$. If the
    universe expands isentropically $Y_{\j}$ is not affected by the
    expansion of the universe, so that $Y_{\j}$ can only be changed by 
    interactions. 

    We can distinguish between elastic and inelastic scatterings.
    Elastic scatterings only affect the phase space densities of the
    particles but not the number densities, whilst inelastic
    scatterings do change the number densities. If elastic scatterings
    do occur at a higher rate than inelastic scatterings we can assume
    kinetic equilibrium, so that the phase space density is \cite{kw}
    \beq
     f_{\j}(E_{\j},T)={n_{\j}\over n_{\j}^{eq}}\mbox{e}^{-E_{\j}/T}\;.
    \eeq
    Consequently the Boltzmann equation describing the evolution of 
    $Y_{\j}$ is (cf.\ \cite{kw,luty})
    \beqa
    {\mbox{d}Y_{\j}\over\mbox{d}z}&=&-{z\over sH\left(m_{\j}\right)}
    \sum\limits_{a,i,j,\ldots}\left[{Y_{\j}Y_a\ldots\over
     Y_{\j}^{eq}Y_a^{eq}\ldots}\,\g^{eq}\left(\j+a+\ldots\to 
     i+j+\ldots\right)-\right.\NO\\[1ex]
     &&\qquad\qquad\qquad\left.-{Y_iY_j\ldots\over 
     Y_i^{eq}Y_j^{eq}\ldots}
     \,\g^{eq}\left(i+j+\ldots\to\j+a+\ldots\right)\right]\;,
    \label{7}
    \eeqa
    where $z=m_{\j}/T$ and $H\left(m_{\j}\right)$ is the Hubble
    parameter at $T=m_{\j}$. 

    The right-hand side of eq.~(\ref{7}) describes the
    interactions in which a $\j$ particle takes part, where $\g^{eq}$
    is the space time density of scatterings in thermal equilibrium. 
    In a dilute gas we only have to take into account decays,
    two-particle scatterings and the corresponding back reactions. For
    a decay one finds \cite{luty}
    \beq
     \g_D:=\g^{eq}(\j\to i+j+\ldots)=
     n^{eq}_{\j}{\mbox{K}_1(z)\over\mbox{K}_2(z)}\,\tilde{\G}_{rs}\;,
    \eeq
    where K$_1$ and K$_2$ are modified Bessel functions and
    $\tilde{\G}_{rs}$ is the usual decay width in the rest system of
    the decaying particle. The ratio of the Bessel functions is a
    time dilatation factor. 

    If we neglect $CP$ violating effects we have the same reaction
    density for inverse decays,
    \beq
     \g_{ID}:=\g^{eq}(i+j+\ldots\to\j)=\g_D\;.
    \eeq

    For two body scattering one has
    \beq
     \g^{eq}({\j}+a\leftrightarrow i+j+\ldots)=
     {T\over64\p^4}\int\limits_{\left(m_{\j}+m_a\right)^2}^{\infty}
     \hspace{-0.5cm}\dd s\,\hat{\s}(s)\,\sqrt{s}\,
     \mbox{K}_1\left({\sqrt{s}\over T}\right)\;,
    \eeq 
    where $s$ is the squared center of mass energy and the reduced
    cross section $\hat{\s}(s)$ for the process ${\j}+a\to i+j+\ldots$
    is related to the usual total cross section $\s(s)$ by
    \beq
      \hat{\s}(s)={8\over s}
      \left[\left(p_{\j}\cdot p_a\right)^2-m_{\j}^2m_a^2\right]\,\s(s)\;.
    \eeq

    Since we have assumed kinetic equilibrium, contributions from
    elastic scatterings drop out of eq.~(\ref{7}). Hence we only have
    to take into account inelastic processes.
\section{The model \label{model}}
\subsection{Gauge and Yukawa couplings}
    The 16 plet of SO$(10)$ contains, in addition to the 15 Weyl
    fermions of one standard model quark-lepton family, a right-handed
    neutrino $\n_R$ which is a singlet under the standard model gauge
    group.  It is, therefore, natural to embed the baryogenesis
    mechanism of Fukugita and Yanagida into a SO$(10)$ GUT.

    To explain the unification of the coupling constants in SO$(10)$
    one needs an intermediate breaking scale $v'$ of the order of
    $10^{10}$ to $10^{13}\,$GeV (cf.\ \cite{fy1}). The intermediate
    symmetry could be a left-right-symmetry or a Pati-Salam-symmetry.
    For simplicity we take the minimal extension of the standard model
    which is based on the gauge group
    \beq
     G=\mbox{SU}(3)_C\times\mbox{SU}(2)_L\times\mbox{U}(1)_Y\times
     \mbox{U}(1)_{Y'}\;.
    \eeq
    This gauge group cannot explain the gauge coupling unification
    \cite{fy1} but it may be regarded as a toy model for the other
    symmetry groups. We then have the following breaking scheme
    \beqa
       \mbox{SO}(10)\to\qquad\ldots&\longrightarrow&\mbox{SU}(3)_C\times
       \mbox{SU}(2)_L\times\mbox{U}(1)_Y\times\mbox{U}(1)_{Y'}\NO\\
       &\stackrel{\displaystyle \langle\chi\rangle=v'}{\longrightarrow}&
       \mbox{SU}(3)_C\times\mbox{SU}(2)_L\times\mbox{U}(1)_Y\NO\\
       &\stackrel{\displaystyle \langle\phi\rangle=v}{\longrightarrow}&
       \mbox{SU}(3)_C\times\mbox{U}(1)_{em}\;.\NO
    \eeqa
    Here $\f=(\vf^0,\vf^{-})$ is the standard model Higgs doublet and
    $\c$ is the Higgs boson needed for the breaking of the
    U$(1)_{Y'}$. The
    $\mbox{SU}(2)_L\times\mbox{U}(1)_Y\times\mbox{U}(1)_{Y'}$ part of
    the lagrangian is \cite{b2}
    \beqa
      \cl&=&-{1\over4}\vec{W}_{\m\n}\vec{W}^{\m\n}
      -{1\over4}B_{\m\n}B^{\m\n}-{1\over4}C_{\m\n}C^{\m\n}\NO\\[1ex]
      &&+i\,\overline{l_L}\,\Slash{D}\,l_L+i\,\overline{\n_R}\,
      \Slash{D}\,\n_R+i\,\overline{e_R}\,\Slash{D}\,e_R
      +\left(D_{\m}\f\right)^{\dg}\left(D^{\m}\f\right)+
      \left(D_{\m}\c\right)^{\dg}\left(D^{\m}\c\right)\NO\\[1ex]
      &&-\left(\overline{l_L}\,\f\,g_{\n}\,\n_R
      +\overline{l_L}\,\tilde{\f}\,g_e\,e_R
      +{1\over2}\,\c\,\overline{\n^C_R}\,h\,\n_R
      +\mbox{ h.c.}\right)\;,
    \eeqa
    where we have omitted the quark fields. $l_L=\left(\n_L,e_L\right)$ 
    is the left-handed lepton doublet and $\n_R$ is the
    right-handed neutrino. The charge conjugated field $\n^C_R$ is
    defined by $\n^C_R=C\overline{\n_R}^T$, where $C$ is the charge
    conjugation matrix. $g_e$, $g_{\n}$ and $h$ are the Yukawa
    coupling matrices which are responsible for the lepton masses. One
    can always choose the lepton fields in such a way that $g_e$ and
    $h$ are diagonal and real. 
    
    The covariant derivative has the form
    \beq
      D_{\m}=\partial_{\m}-ig\vec{W}_{\m}\cdot\vec{T}-ig'B_{\m}Y
      -ig'\sqrt{2\over3}C_{\m}Y'\label{2_5}\;,
    \eeq
    where $\vec{W}^{\m}$, $B^{\m}$ and $C^{\m}$ are the SU$(2)_L$,
    U$(1)_Y$ and U$(1)_{Y'}$ gauge fields. $\vec{W}^{\m\n}$,
    $B^{\m\n}$ and $C^{\m\n}$ are the corresponding field strength
    tensors. Because of their common origin in the SO$(10)$ both
    abelian groups have the same gauge coupling constant $g'$. From
    the structure of SO$(10)$ it follows that
    $Y'=Y-{5\over4}(B-L)$. Therefore the fermions and the Higgs bosons
    carry the following $Y'$-charges
    \beq
     \begin{array}{l@{\qquad}l@{\qquad}l}
     \displaystyle Y'\left(l_L\right)={3\over4}&
     \displaystyle Y'\left(e_R\right)={1\over4}&
     \displaystyle Y'\left(\n_R\right)={5\over4}\\[2ex]
     \displaystyle Y'\left(q_L\right)=-{1\over4}&
     \displaystyle Y'\left(u_R\right)={1\over4}&
     \displaystyle Y'\left(d_R\right)=-{3\over4}\\[2ex]
     \displaystyle Y'\left(\f\right)=-{1\over2}&
     \displaystyle Y'\left(\c\right)=-{5\over2}\;.&
     \end{array}
    \eeq
    The most general Higgs potential is
    \beq
      V(\c,\f)=\m_1\,\f^{\dg}\f+\m_2\,\c^{\dg}\c+
      {1\over2}\l_1\left(\f^{\dg}\f\right)^2+
      {1\over2}\l_2\left(\c^{\dg}\c\right)^2+
      \l_3\left(\c^{\dg}\c\right)\left(\f^{\dg}\f\right)\;,
    \eeq
    where $\l_3>-\sqrt{\l_1\l_2}$, so that the potential is bounded
    from below.  Spontaneous symmetry breaking leads to two massive
    neutral gauge bosons $Z$ and $Z'$.

    The right-handed neutrinos acquire Majorana masses $M=hv'$ when
    the U$(1)_{Y'}$ is broken. Additionally the neutrinos get Dirac
    masses $m_D=vg_{\n}$ with $v=174\,$GeV at the electroweak symmetry
    breaking. This offers the possibility to explain the smallness of
    the $\n_e$, $\n_{\m}$ and $\n_{\t}$ masses via the seesaw
    mechanism \cite{seesaw}.

    Since we want to explain the generation of a lepton asymmetry
    before the electroweak phase transition, we can take
    $\VEV{\f}=v=0$, so that all mixing angles vanish. Therefore, all
    the standard model particles are massless and the
    additional neutral gauge boson $Z'$ is identical to the vector
    field $C^{\m}$. For $\VEV{\c}=v'\ne0$ the mass of the $Z'$ is
    \beq
     m_{Z'}={5\over\sqrt{3}}\,g'\,v'\;.
    \eeq
    Since Yukawa couplings are usually small when compared to gauge
    couplings, one expects that the Majorana masses $M_i$ of the
    right-handed neutrinos are small compared to the $Z'$ mass.

    The Higgs boson $H_0'$ contained in the field $\c$ gets the
    mass 
    \beq
     m_{H_0'}=\sqrt{\l_2}\,v'\;.
    \eeq 
    To realize the assumed hierarchy $v'\gg v$ one needs
    $\l_2\ge2\cdot10^{-3}$ \cite{en}. Hence $m_{H_0'}$ is of
    the order of $m_{Z'}$, so that processes in which a $H_0'$
    takes part are kinematically suppressed at temperatures
    below $m_{Z'}$. Moreover the Yukawa couplings $h$ of $H_0'$ were
    assumed to be small in comparison with the gauge couplings.
    Therefore the Higgs boson $H_0'$ can be neglected in the
    following.

    Since we have assumed that the matrix $h$ is real and diagonal the
    weak eigenstates are equal to the Majorana mass eigenstates,
    \beq
    \n=\n_L+\n_L^C\qquad\mbox{and}\qquad N=\n_R+\n_R^C\;,
    \eeq
    where $\n$ and $N$ are four-component Majorana spinors.
\subsection{Reaction rates for lepton number violating processes}
    We are now able to calculate the relevant reaction rates. It
    appears reasonable to assume a mass hierarchy of the form
    $M_1\ll M_2\ll M_3$ for the right-handed neutrinos, so that the
    lightest right-handed neutrino, $N^1$, will still be in
    equilibrium when $N^2$ and $N^3$ decay. The lepton number
    violating interactions mediated by $N^1$ will, therefore,
    wash out an asymmetry generated by the decays of the neutrinos
    $N^2$ and $N^3$. Hence a significant lepton asymmetry can only be
    generated by $N^1$-decays and we expect that the temperature at
    which the asymmetry is generated is of the order of $M_1$.
    Consequently it is useful to relate all the masses and energies to
    $M_1$. We define the following dimensionless quantities
    \beq
     a_i:={M_i^2\over M_1^2}\;,\qquad\qquad
     y := {m_{Z'}^2\over M_1^2}\qquad\mbox{and}\qquad
     x:={s\over M_1^2}\;.
    \eeq
    \begin{figure}[t]
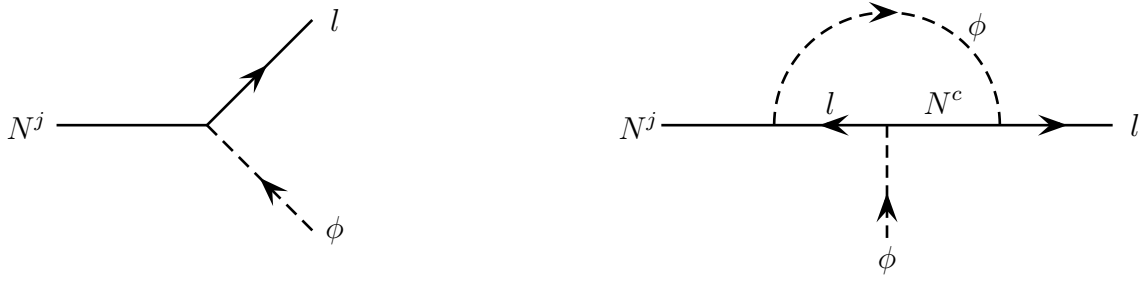

     \begin{center}
     \pspicture(1,0.3)(8,3.5)
     \psline[linewidth=1pt](1.6,2)(3.6,2)
     \rput[cc]{0}(1.2,2){$\displaystyle N^j$}
     \psline[linewidth=1pt](3.6,2)(5,3.4)
     \psline[linewidth=2pt]{->}(4.3,2.7)(4.4,2.8)
     \rput[cc]{0}(5.3,3.4){$\displaystyle l$}
     \psline[linewidth=1pt,linestyle=dashed](3.6,2)(5,0.6)
     \psline[linewidth=2pt]{->}(4.4,1.2)(4.3,1.3)
     \rput[cc]{0}(5.3,0.6){$\displaystyle \f$}
     \endpspicture
     \hspace{1cm}
     \pspicture(0.5,0.3)(7.5,3.5)
     \psline[linewidth=1pt](1,2)(7,2)
     \rput[cc]{0}(0.7,2){$\displaystyle N^j$}
     \rput[cc]{0}(3.25,2.3){$\displaystyle l$}
     \rput[cc]{0}(7.3,2){$\displaystyle l$}
     \rput[cc]{0}(4.75,2.3){$\displaystyle N^c$}
     \psline[linewidth=2pt]{->}(3.2,2)(3.1,2)
     \psline[linewidth=2pt]{->}(6.3,2)(6.4,2)
     \psarc[linewidth=1pt,linestyle=dashed](4,2){1.5}{0}{180}
     \psline[linewidth=2pt]{->}(4.05,3.5)(4.15,3.5)
     \rput[cc]{0}(5.2,3.3){$\f$}
     \psline[linewidth=1pt,linestyle=dashed](4,2)(4,0.5)
     \psline[linewidth=2pt]{->}(4,1)(4,1.1)
     \rput[cc]{0}(4,0.2){$\f$}
     \endpspicture
     \caption{\it Right-handed neutrino decay at tree and one-loop level.
       \label{dec}}
     \end{center}
    \end{figure}
    The lepton asymmetry is generated by the $CP$ violating decay of
    the right-handed neutrinos. Making use of the relation
    \beq
    g_{\n}=m_D{1\over v}=m_D{g\over\sqrt{2}M_W}\;,
    \eeq
    where $M_W=80\,$GeV is the $W$ boson mass and $g$ is the SU$(2)$
    coupling constant, one finds for the decay width at tree level 
    \cite{fy2}
    \beq
      \tilde{\G}_{Dj}:=\tilde{\G}_{rs}\left(N^j\to\f^{\dg}+l\right)+
      \tilde{\G}_{rs}\left(N^j\to\f+\overline{l}\right)={\a\over\sin^2\q}
      {M_j\over4}{(m_D^{\dg}m_D)_{jj}\over M_W^2}\;.
    \eeq
    Here $\q$ is the weak mixing angle. The leading contribution to
    the $CP$-asymmetry in the decay of $N^j$ reads 
    \beq
     \ve_j:={\tilde{\G}_{rs}\left(N^j\to\f^{\dg}+l\right)-
     \tilde{\G}_{rs}\left(N^j\to\f+\overline{l}\right)\over
     \tilde{\G}_{rs}\left(N^j\to\f^{\dg}+l\right)+
     \tilde{\G}_{rs}\left(N^j\to\f+\overline{l}\right)}
     \label{30}\;.
    \eeq
    It is due to the interference between the tree
    level decay amplitude and the one loop amplitude shown in
    fig.~\ref{dec}. One finds\footnote{The results quoted in the
      literature differ from our result by a factor $8$ 
      (ref.~\cite{luty}), $2$ (ref.~\cite{fy1}) or $18$ (ref.~\cite{fy2})}
    \beqa
     &&\ve_j={\a\over 4M_W^2\sin^2\q}{1\over(m_D^{\dg}m_D)_{jj}}
     \sum\limits_c\mbox{Im}\left[\left(m_D^{\dg}m_D\right)_{jc}^2\right]\,
     f\left({a_c\over a_j}\right)\\[1ex]
     &&\mbox{with}\quad f(x)=\sqrt{x}\left[1-(1+x)\ln\left({1+x\over x}
     \right)\right]\;.
    \eeqa
    From the definition (\ref{30}) of the $CP$ asymmetry it follows
    that the reaction densities for decays and inverse decays can be 
    parametrized in the following way
    \beqa
     \g^{eq}\left(N^j\to\f^{\dg}+l\right)=\g^{eq}\left(\f+\overline{l}\to 
     N^j\right)&=&{1\over2}(1+\ve_j)\,\g_{Dj}\\[1ex]
     \g^{eq}\left(N^j\to\f+\overline{l}\right)=\g^{eq}\left(\f^{\dg}+l\to
     N^j\right)&=&{1\over2}(1-\ve_j)\,\g_{Dj}\\[1ex]
     \mbox{with}\qquad\g_{Dj}:=\g^{eq}\left(N^j\to\f^{\dg}+l\right)+
     \g^{eq}\left(N^j\to\f+\overline{l}\right)&=&\tilde{\G}_{Dj}\,
     n_{N^j}^{eq}\,{\mbox{K}_1(M_j/T)\over\mbox{K}_2(M_j/T)}\;.
    \eeqa
    \begin{figure}[t]
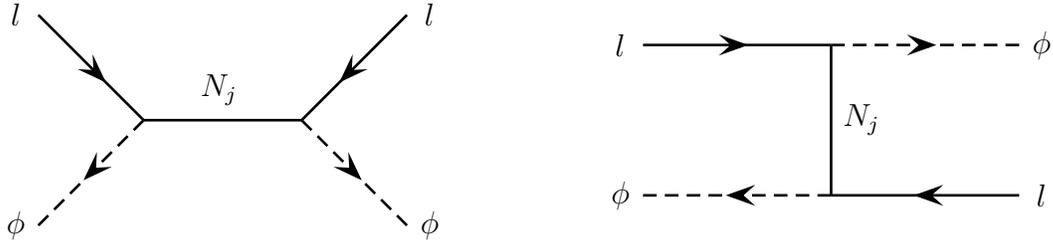

     \begin{center}
     \pspicture[0.5](1,0.3)(8,3.5)
     \psline[linewidth=1pt,linestyle=dashed](1.6,0.6)(3,2)
     \psline[linewidth=2pt]{<-}(2.2,1.2)(2.3,1.3)
     \rput[cc]{0}(1.3,0.6){$\displaystyle \f$}
     \psline[linewidth=1pt](1.6,3.4)(3,2)
     \psline[linewidth=2pt]{->}(2.4,2.6)(2.5,2.5)
     \rput[cc]{0}(1.3,3.4){$\displaystyle l$}
     \psline[linewidth=1pt](3,2)(5.1,2)
     \rput[cc]{0}(4,2.4){$\displaystyle N_j$}
     \psline[linewidth=1pt](5.1,2)(6.5,3.4)
     \psline[linewidth=2pt]{->}(5.7,2.6)(5.6,2.5)
     \rput[cc]{0}(6.8,3.4){$\displaystyle l$}
     \psline[linewidth=1pt,linestyle=dashed](5.1,2)(6.5,0.6)
     \psline[linewidth=2pt]{->}(5.8,1.3)(5.9,1.2)
     \rput[cc]{0}(6.8,0.6){$\displaystyle \f$}
     \endpspicture
     \hspace{1cm}
     \pspicture[0.5](8.5,0.3)(15,3.5)
     \psline[linewidth=1pt](9,3)(11.5,3)
     \psline[linewidth=2pt]{->}(10.3,3)(10.4,3)
     \rput[cc]{0}(8.7,3){$\displaystyle l$}
     \psline[linewidth=1pt,linestyle=dashed](11.5,3)(14,3)
     \psline[linewidth=2pt]{->}(12.8,3)(12.9,3)
     \rput[cc]{0}(14.3,3){$\displaystyle \f$}
     \psline[linewidth=1pt](11.5,3)(11.5,1)
     \rput[cc]{0}(11.9,2){$\displaystyle N_j$}
     \psline[linewidth=1pt,linestyle=dashed](9,1)(11.5,1)
     \psline[linewidth=2pt]{<-}(10.1,1)(10.2,1)
     \rput[cc]{0}(8.7,1){$\displaystyle \f$}
     \psline[linewidth=1pt](11.5,1)(14,1)
     \psline[linewidth=2pt]{<-}(12.6,1)(12.7,1)
     \rput[cc]{0}(14.3,1){$\displaystyle l$}
     \endpspicture
     \caption{\it Lepton number violating lepton Higgs scattering
       \label{lept}}
     \end{center}
    \end{figure}
    Since the $CP$-asymmetry is a higher order effect we have to take
    into account other higher order effects as well, especially the
    $L$ violating lepton Higgs scattering $l+\f^{\dg}\to\overline{l}+\f$ 
    shown in fig.~\ref{lept} with a right-handed neutrino as intermediate
    state. As shown in ref.~\cite{kw} this diagram is necessary to
    avoid the generation of a lepton asymmetry in thermal equilibrium.

    However, the scatterings with a real intermediate neutrino were
    already taken into account, because they can be described by an
    inverse decay followed by a decay. Therefore, we have to remove the
    contributions from physical intermediate states from the reduced
    cross section\cite{kw}.

    In \cite{luty} the flavour structure of this diagram was
    neglected, i.e.\ only the contribution from the lightest
    right-handed neutrino $N^1$ was taken into account. We will take
    into consideration two right-handed neutrinos, $N^1$ and $N^2$.
    When discussing our results we will see that the third neutrino can
    indeed be neglected. The reduced cross section is then given by:
    \beqa
     \hspace{-2pt}\hat{\s}'_N(s)&\hspace{-8pt}=&\hspace{-8pt}
     {\a^2\over\sin^4\q}{2\p\over M_W^4}{1\over x}
     \left\{\sum\limits_{j=1}^2 a_j\left(m_D^{\dg}m_D\right)^2_{jj}
     \left[{x\over a_j}+{2x\over D_j(x)}+{x^2\over2D_j^2(x)}-
     \left(1+2{x+a_j\over D_j(x)}\right)\ln\left({x+a_j\over a_j}
     \right)\right]\right.\NO\\[1ex]
     &&+2\sqrt{a_1a_2}\,\mbox{Re}\left[\left(m_D^{\dg}m_D\right)^2_{12}
     \right]\left[{x\over D_1(x)}+{x\over D_2(x)}+{x^2\over2D_1(x)D_2(x)}
     \right.\NO\\[1ex]
     &&\left.\left.-{(x+a_1)(x+a_1-2a_2)\over D_2(x)(a_1-a_2)}
     \ln\left({x+a_1\over a_1}\right)-{(x+a_2)(x+a_2-2a_1)\over 
     D_1(x)(a_2-a_1)}\ln\left({x+a_2\over a_2}\right)\right]\right\}\;.
    \eeqa
    The prime denotes that we have subtracted the contributions from
    real intermediate states and
    \beq
    {1\over D_j(x)}:={x-a_j\over (x-a_j)^2+a_jc_j}\;,\qquad
    \mbox{with }c_j:=\left({\tilde{\G}_{Dj}\over M_1}\right)^2\;,
    \eeq
    is the off-shell part of the propagator.

    There are some other $L$ violating processes mediated by a
    right-handed neutrino in the $t$-channel, like the scattering 
    $l+l\to\f+\f$. The reduced cross section for this process, which
    was neglected in \cite{luty}, is
    \beqa
     \lefteqn{\hat{\s}_{N,t}(s)={2\p\a^2\over M_W^4\sin^4\q}\left\{
     \sum\limits_{j=1}^2a_j\left(m_D^{\dg}m_D\right)_{jj}^2\left[
     {x\over x+a_j}+{1\over x+2a_j}\ln\left({x+a_j\over a_j}\right)\right]
     \right.}\\[1ex]
     &+&\hspace{-4pt}\left.\mbox{Re}\left[\left(m_D^{\dg}
     m_D\right)_{12}^2\right]{\sqrt{a_1a_2}\over(a_1-a_2)(x+a_1+a_2)}
     \left[(x+2a_1)\ln\left({x+a_2\over a_2}\right)-
     (x+2a_2)\ln\left({x+a_1\over a_1}\right)\right]\right\}\;.\NO
    \eeqa
    The same result is valid for the process 
    $\overline{l}+\overline{l}\to\f^{\dg}+\f^{\dg}$ and the back
    reactions.

    \begin{figure}
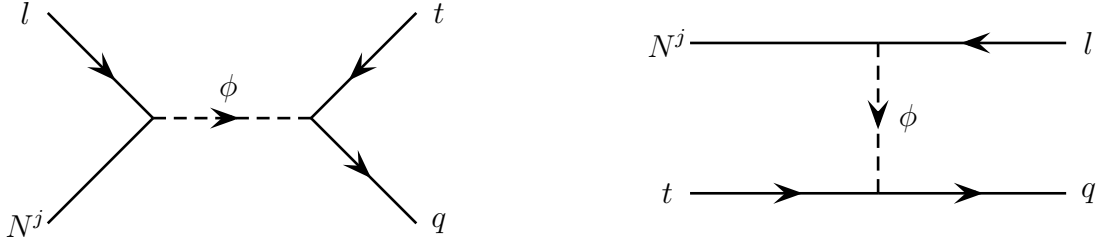

      \begin{center}
     \pspicture[0.5](0.5,0.3)(8.5,3.7)
     \psline[linewidth=1pt](1.6,0.6)(3,2)
     \rput[cc]{0}(1.3,0.6){$\displaystyle N^j$}
     \psline[linewidth=1pt](1.6,3.4)(3,2)
     \psline[linewidth=2pt]{->}(2.4,2.6)(2.5,2.5)
     \rput[cc]{0}(1.3,3.4){$\displaystyle l$}
     \psline[linewidth=1pt,linestyle=dashed](3,2)(5.1,2)
     \psline[linewidth=2pt]{->}(4.05,2)(4.15,2)
     \rput[cc]{0}(4,2.4){$\displaystyle \f$}
     \psline[linewidth=1pt](5.1,2)(6.5,3.4)
     \psline[linewidth=2pt]{->}(5.7,2.6)(5.6,2.5)
     \rput[cc]{0}(6.8,3.4){$\displaystyle t$}
     \psline[linewidth=1pt](5.1,2)(6.5,0.6)
     \psline[linewidth=2pt]{->}(5.8,1.3)(5.9,1.2)
     \rput[cc]{0}(6.8,0.6){$\displaystyle q$}
     \endpspicture
     \hspace{0.5cm}
     \pspicture[0.5](8,0.3)(16,3.7)
     \psline[linewidth=1pt](9,3)(11.5,3)
     \rput[cc]{0}(8.7,3){$\displaystyle N^j$}
     \psline[linewidth=1pt](11.5,3)(14,3)
     \psline[linewidth=2pt]{<-}(12.6,3)(12.7,3)
     \rput[cc]{0}(14.3,3){$\displaystyle l$}
     \psline[linewidth=1pt,linestyle=dashed](11.5,3)(11.5,1)
     \psline[linewidth=2pt]{->}(11.5,1.95)(11.5,1.85)
     \rput[cc]{0}(11.9,2){$\displaystyle \f$}
     \psline[linewidth=1pt](9,1)(11.5,1)
     \psline[linewidth=2pt]{->}(10.3,1)(10.5,1)
     \rput[cc]{0}(8.7,1){$\displaystyle t$}
     \psline[linewidth=1pt](11.5,1)(14,1)
     \psline[linewidth=2pt]{->}(12.8,1)(12.9,1)
     \rput[cc]{0}(14.3,1){$\displaystyle q$}
     \endpspicture
     \end{center}
     \caption{\it Lepton number violating scatterings mediated by a
       standard model Higgs boson in the $s$- or $t$-channel
       \label{top}} 
    \end{figure}
    The Yukawa couplings of the standard model are very small and can
    be neglected, with the exception of the term responsible for the
    top quark mass $m_t$. Thus we have to take into account the $L$
    violating interactions of a right-handed neutrino with a
    top-quark. There are $s$- and $t$-channel contributions, shown in 
    fig.~\ref{top}. The reduced cross section for the $s$-channel 
    process $N^j+l\to\overline{t}+q$ is \cite{luty}
    \beq
     \hat{\s}_{\f,s}^j(s)={3\p\a^2m_t^2\over M_W^4\sin^4\q}\left(m_D^{\dg}m_D
     \right)_{jj}\left({x-a_j\over x}\right)^2\;.
    \eeq
    One finds the same result for the process $N^j+\overline{l}\to
    t+\overline{q}$ and the corresponding back reactions.

    To get a well defined result for the $t$-channel contribution one
    has to introduce a Higgs mass $m_{\f}$. Keeping only the leading
    contributions for $m_{\f}\to0$ one finds \cite{luty}
    \beq
     \hat{\s}_{\f,t}^j(s)={3\p\a^2m_t^2\over M_W^4\sin^4\q}\left(m_D^{\dg}m_D
     \right)_{jj}\left[{x-a_j\over x}+{a_j\over x}
     \ln\left({x-a_j+y'\over y'}\right)\right]\;,
    \eeq
    with the dimensionless squared Higgs mass $y':=m_{\f}^2/M_1^2.$
    Since these processes have only a small influence on the generated
    asymmetry, the results are very insensitive to the value of
    $m_{\f}$. In the following we use $m_{\f}=800\,$GeV.
\subsection{Reaction rates for lepton number conserving processes}
    \begin{figure}[t]
    \begin{center}
     \pspicture(1,0)(15,4)
     \rput[cc]{0}(4,4){(a)}
     \psline[linewidth=1pt](1.6,0.6)(3,2)
     \psline[linewidth=2pt]{<-}(2.2,1.2)(2.3,1.3)
     \rput[cc]{0}(1.1,0.6){$\displaystyle f,\f$}
     \psline[linewidth=1pt](1.6,3.4)(3,2)
     \psline[linewidth=2pt]{->}(2.4,2.6)(2.5,2.5)
     \rput[cc]{0}(1.1,3.4){$\displaystyle f,\f$}
     \pscoil[coilarm=0.1cm,linewidth=1pt,coilwidth=0.3cm,coilaspect=0]
            (3,2)(5.1,2)
     \rput[cc]{0}(4,2.4){$\displaystyle Z'$}
     \psline[linewidth=1pt](5.1,2)(6.5,3.4)
     \rput[cc]{0}(6.8,3.4){$\displaystyle N^j$}
     \psline[linewidth=1pt](5.1,2)(6.5,0.6)
     \rput[cc]{0}(6.8,0.6){$\displaystyle N^j$}
     \rput[cc]{0}(12,4){(b)}
     \psline[linewidth=1pt](9.6,0.6)(11,2)
     \rput[cc]{0}(9.3,0.6){$\displaystyle N^i$}
     \psline[linewidth=1pt](9.6,3.4)(11,2)
     \rput[cc]{0}(9.3,3.4){$\displaystyle N^i$}
     \pscoil[coilarm=0.1cm,linewidth=1pt,coilwidth=0.3cm,coilaspect=0]
            (11,2)(13.1,2)
     \rput[cc]{0}(12,2.4){$\displaystyle Z'$}
     \psline[linewidth=1pt](13.1,2)(14.5,3.4)
     \rput[cc]{0}(14.8,3.4){$\displaystyle N^j$}
     \psline[linewidth=1pt](13.1,2)(14.5,0.6)
     \rput[cc]{0}(14.8,0.6){$\displaystyle N^j$}
     \endpspicture\\\mbox{ }\\
     \pspicture(1,1)(15,3.5)
     \rput[cc]{0}(4,4){(c)}
     \psline[linewidth=1pt](1.6,0.6)(3,2)
     \rput[cc]{0}(1.3,0.6){$\displaystyle N^i$}
     \psline[linewidth=1pt](1.6,3.4)(3,2)
     \rput[cc]{0}(1.3,3.4){$\displaystyle N^i$}
     \pscoil[coilarm=0.1cm,linewidth=1pt,coilwidth=0.3cm,coilaspect=0]
            (3,2)(5.1,2)
     \rput[cc]{0}(4,2.4){$\displaystyle Z'$}
     \psline[linewidth=1pt](5.1,2)(6.5,3.4)
     \rput[cc]{0}(6.8,3.4){$\displaystyle N^i$}
     \psline[linewidth=1pt](5.1,2)(6.5,0.6)
     \rput[cc]{0}(6.8,0.6){$\displaystyle N^i$}
     \rput[cc]{0}(12,4){(d)}
     \psline[linewidth=1pt](9.5,3)(14.5,3)
     \rput[cc]{0}(9.2,3){$\displaystyle N^i$}
     \rput[cc]{0}(14.8,3){$\displaystyle N^i$}
     \pscoil[coilarm=0.1cm,linewidth=1pt,coilwidth=0.3cm,coilaspect=0]
            (12,3)(12,0.9)
     \rput[cc]{0}(12.4,1.95){$\displaystyle Z'$}
     \psline[linewidth=1pt](9.5,0.9)(14.5,0.9)
     \rput[cc]{0}(9.2,0.9){$\displaystyle N^i$}
     \rput[cc]{0}(14.8,0.9){$\displaystyle N^i$}
     \endpspicture
    \end{center}
     \caption[Lepton number conserving processes]{\begin{tabular}[t]{ll}
          \multicolumn{2}{l}{\it Lepton number conserving processes:}\\
         (a)&\it $N^j$ pair creation\\
         (b)&\it process $N^i+N^i\to N^j+N^j$ $(i\ne j)$\\
         (c,d)&\it elastic scattering $N^i+N^i\to N^i+N^i$
    \label{Lcons}\end{tabular}}
    \end{figure}
    The lepton number violating processes are too weak to bring the
    right-handed neutrinos into equilibrium at high temperatures.
    Therefore, one has to consider lepton number conserving processes,
    which were neglected up to now, like the $N^j$ pair creation and
    pair annihilation. The dominating contribution comes from the
    $Z'$ exchange in the $s$-channel shown in fig.~\ref{Lcons}a.
    The total reduced cross section for the processes
    $f+\overline{f}\leftrightarrow N^j+N^j$, where $f$ is a standard
    model fermion, and $\f+\f^{\dag}\leftrightarrow N^j+N^j$ is
    \beq
     \hat{\s}_{Z'}(s)={4225\p\over216}\,{\a^2\over\cos^4\q}\,
     {\sqrt{x}\over(x-y)^2+yc}\,\left(x-4a_j\right)^{3/2}\;.
    \eeq 
    To get a well defined result we have introduced the $Z'$ decay
    width into the propagator,
    \beq
    c:=\left({\tilde{\G}_{Z'}\over M_1}\right)^2\qquad\mbox{with}\qquad
    \tilde{\G}_{Z'}={\a m_{Z'}\over\cos^2\q}\,\left[{25\over18}
    \sum\limits_i\left({y-4a_i\over4y}\right)^{3/2}\,
    \q\left(y-4a_i\right)+{169\over144}\right]\;.
    \eeq

    Since we will have different heavy neutrino flavours we have
    to consider transitions between right-handed neutrinos like
    $N^i+N^i\leftrightarrow N^j+N^j$ $(i\ne j)$ with an intermediate 
    $Z'$ in the $s$-channel, shown in fig.~\ref{Lcons}b. The reduced
    cross section for this process is
    \beq
     \hat{\s}_{N^iN^j}(s)={625\p\over216}\,{\a^2\over\cos^4\q}
     \left\{{1\over x}\,{(x-4a_i)^{3/2}\,(x-4a_j)^{3/2}\over 
     (x-y)^2+yc}+12{a_ia_j\over y^2}\sqrt{x-4a_i\over x}
     \sqrt{x-4a_j\over x}\right\}\;.\label{43}
    \eeq

    Finally we have to look at the elastic processes to check that the
    assumed kinetic equilibrium is a good approximation. For this
    purpose we have considered the process $N^i+N^i\to N^i+N^i$. The
    dominating contribution to this transition comes from $Z'$
    exchange in the $s$-channel (fig.~\ref{Lcons}c) and in the $t$-
    and $u$-channel (fig.~\ref{Lcons}d). The reduced cross section for
    this process reads
    \beqa
     \lefteqn{\hat{\s}_{el}(s)={625\p\over72}\,{\a^2\over\cos^4\q}\,
     \left\{{1\over3x}{(x-4a_i)^3\over(x-y)^2+yc}+\right.}\NO\\[1ex]
     &&+{x-4a_i\over xy^2}\,{y(x-4a_i)^2+2(y-2a_i)^3+
     (8a_i^2+3y^2)x-4a_i(y^2+ya_i+4a_i^2)\over x-4a_i+y}+\NO\\[1ex]
     &&+\left.{1\over xy}\,{(3y-4a_i)(x-4a_i)^2+(3x-20a_i)y^2+
     2y(y^2+8a_i^2)\over x-4a_i+2y}
     \ln\left({y\over x-4a_i+y}\right)\right\}\;.
    \eeqa
\section{Results \label{solutions}} 
\subsection{Constraints on the parameters \label{constraints}}
    Before solving the Boltzmann equations we will try to constrain
    the parameters by looking at the reaction rates. The neutrinos
    have to be out of equilibrium when decaying, i.e.\ they have to
    decouple before decaying. Therefore, the decay rate $\G_D$ has to
    be smaller than the Hubble parameter $H$ at temperatures $T\approx
    M_1$. This gives the following constraint \cite{fischler,b3}
    \beq
     \G_{D}(T=M_1)<3H(T=M_1)\qquad\Leftrightarrow\qquad
     \tilde{m}_1:={(m_D^{\dg}m_D)_{11}\over M_1}<9\cdot10^{-3}\,
     \mbox{eV}\;.\label{45}
    \eeq
    $\tilde{m}_1$ is the mass of the lightest neutrino, if $m_D$ is
    approximately diagonal.
    
    Furthermore the pair annihilation rate $\G_{Z'}$ for the neutrinos
    has to be smaller than $3H$ at $T\approx M_1$. This gives a lower
    bound on the $Z'$ mass \cite{luty,en}
    \beq
     \G_{Z'}(T=M_1)<3H(T=M_1)\qquad\Leftrightarrow\qquad
     m_{Z'}>2\cdot10^{11}\,\mbox{GeV}\,\left(M_1\over{10^{10}\,
     \mbox{GeV}}\right)^{3/4}\;.\label{46}
    \eeq
    On the other hand the $L$ asymmetry must be generated before
    the electroweak phase transition, because we need the sphalerons
    which are no longer in equilibrium at temperatures below
    $100\,$GeV. Since the electroweak phase transition takes place
    $10^{-12}\,$s after the big bang, the lifetime of the right-handed
    neutrinos has to be smaller than $10^{-12}\,$s. This yields the
    following constraint
    \beq
     (m_D^{\dg}m_D)_{11}>\left(20\,\mbox{eV}\right)^2
     \left({10^{10}\,\mbox{GeV}\over M_1}\right)\;.
    \eeq
    Finally the reaction rates $\G_N$ and $\G_{N,t}$ for the lepton
    number violating scatterings mediated by the
    right-handed neutrinos must not wash out the generated lepton
    asymmetry at low temperatures. This gives the following condition
    \cite{ht}
    \beq
     \G_N(T=M_1)<3H(T=M_1)\qquad\Leftrightarrow\qquad
     \sum\limits_j\tilde{m}_j^2<\left(7\,\mbox{eV}\right)^2
     \left({10^{10}\,\mbox{GeV}\over M_1}\right)\;.
    \eeq
    The sum over $\tilde{m}_j^2$ can be interpreted as sum over the
    squared masses of the light neutrinos, if $m_D$ is approximately
    diagonal.

\begin{figure}[t]
     \epsfig{file=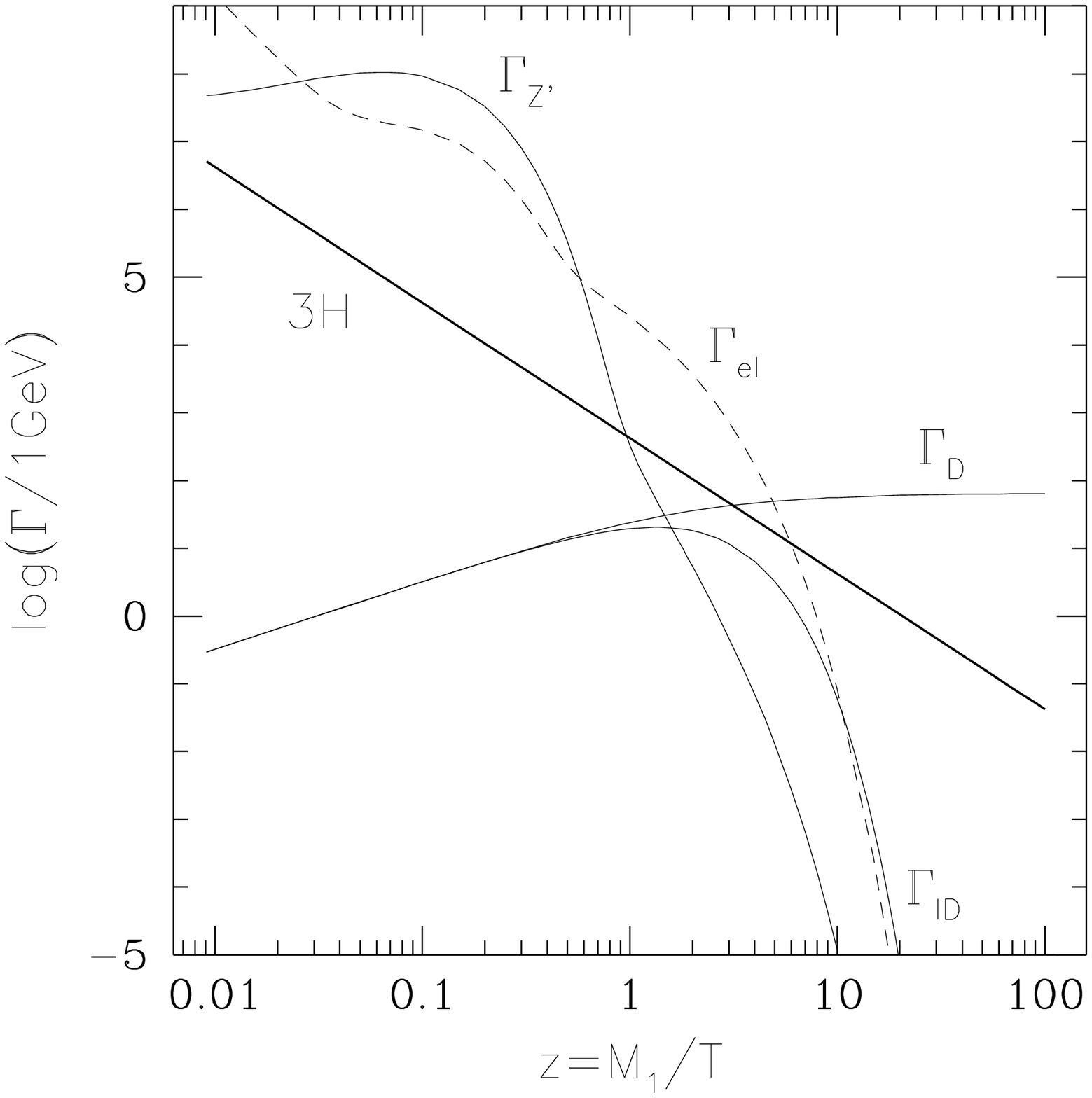,%
             bbllx=44pt,bblly=190pt,%
             bburx=560pt,bbury=690pt,width=8cm}
     \hspace{\fill}
     \epsfig{file=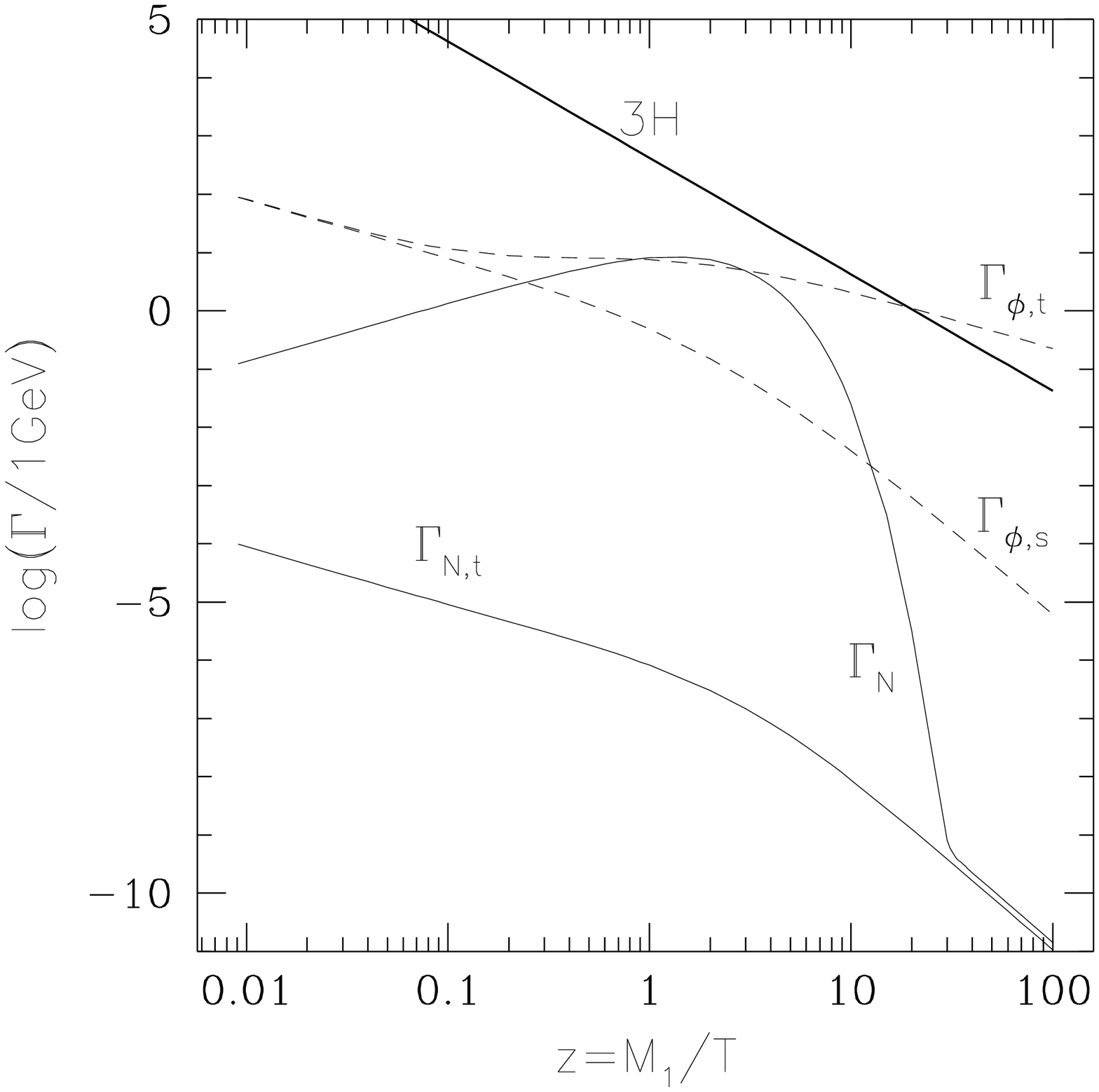,%
             bbllx=44pt,bblly=190pt,%
             bburx=560pt,bbury=690pt,width=8cm}
     \caption{\it Comparison of different reaction rates with the
       Hubble parameter $H$. $\G_{Z'}$ is the pair annihilation rate,
       $\G_D$ the decay rate, $\G_{ID}$ the reaction rate for inverse
       decays, $\G_N$ and $\G_{Nt}$ for lepton number violating
       processes with an intermediate $N^1$, $\G_{\f,s}$ and
       $\G_{\f,t}$ for the interaction of an incoming $N^1$ with a
       top-quark and $\G_{el}$ for the elastic $N^1$-scattering.
       \label{rates}}
    \end{figure}

    We have shown the different reaction rates and the Hubble
    parameter in fig.~\ref{rates}, where we have chosen the 
    following parameters
    \beq
     \tilde{m}_1=10^{-3}\,\mbox{eV}\;,\qquad
     M_1=10^{10}\,\mbox{GeV}\quad\mbox{and}\quad
     m_{Z'}=2\cdot10^{11}\,\mbox{GeV}\;.
    \eeq
    The reaction rates have the correct behaviour. $\G_{Z'}$ is much
    bigger than $3H$ at temperatures $T\gg M_1$, so that the neutrinos
    $N^1$ come into equilibrium at high temperatures, while $\G_{Z'}$
    is small enough at $T\approx M_1$ to avoid the lepton number
    conserving pair annihilation of the neutrinos. Furthermore one can
    see that the reaction rates taken into account in previous
    analyses \cite{fy2,luty}, i.e.\ $\G_D$, $\G_{ID}$, $\G_N$,
    $\G_{\f,s}$ and $\G_{\f,t}$ are far too low to bring the neutrinos
    into thermal equilibrium at high temperatures. 

    Since the decay rate $\G_D$ is smaller than $3H$ for $T\approx
    M_1$, the neutrinos have enough time to deviate from the thermal
    equilibrium, so that the out-of-equilibrium condition is
    fulfilled. Moreover the reaction rates $\G_N$ and
    $\G_{N,t}$ for the lepton number violating interactions and
    $\G_{ID}$ for inverse decays are so small that the $L$ asymmetry
    cannot be erased at temperatures $T<M_1$.

    Finally we have plotted the reaction rate $\G_{el}$ for the
    elastic scatterings $N^1+N^1\leftrightarrow N^1+N^1$. As one can
    see $\G_{el}$ is much bigger than $\G_{Z'}$ in the relevant
    temperature region $T\approx M_1$, so that one can assume kinetic
    equilibrium. 
\subsection{Decay of the lightest heavy neutrino}
\begin{figure}
     \begin{minipage}[t]{8cm}
     \begin{center}\hspace{1cm}(a)\end{center}
     \epsfig{file=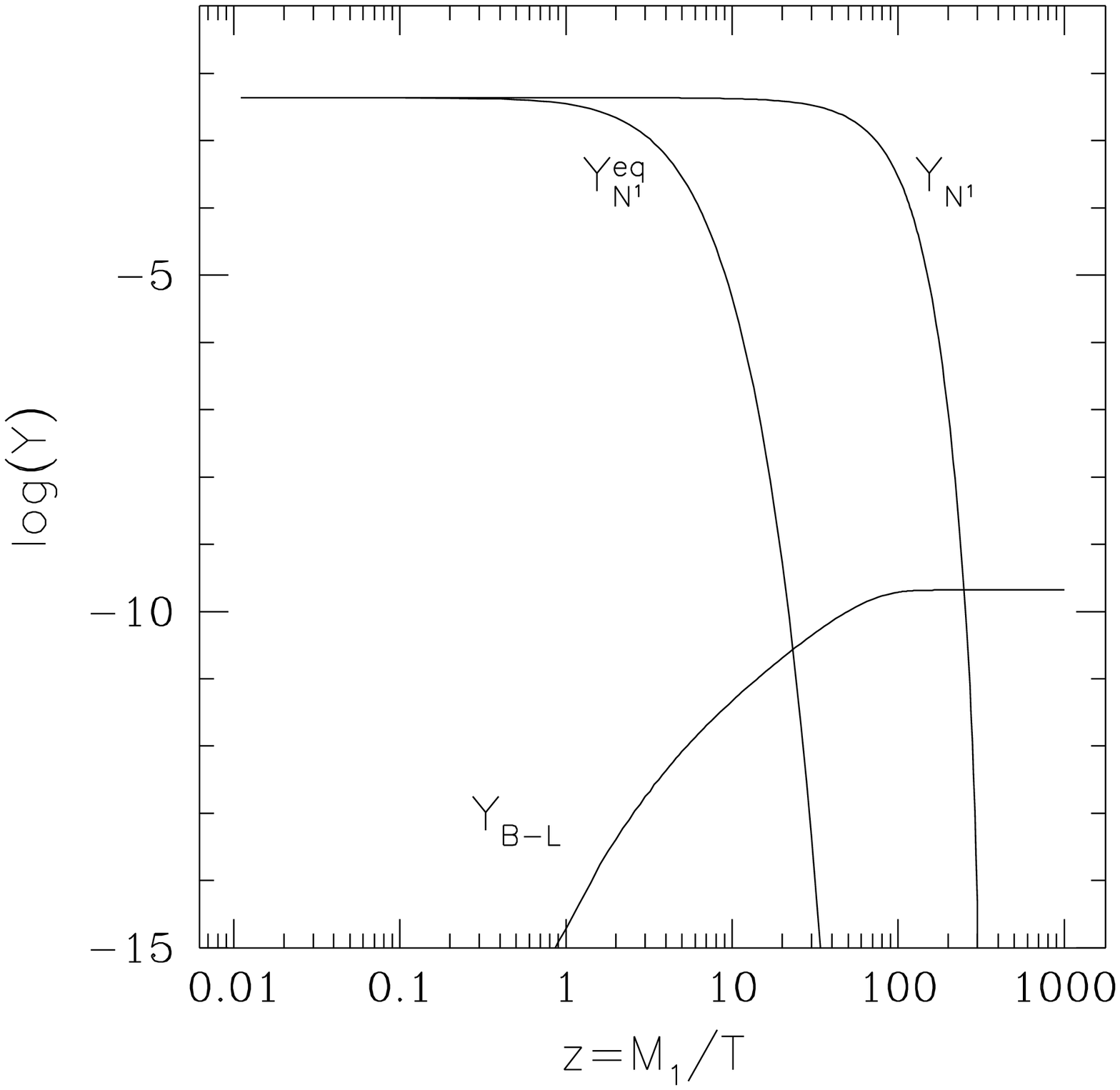,%
             bbllx=44pt,bblly=170pt,%
             bburx=560pt,bbury=690pt,width=8cm}
     \end{minipage}
     \hspace{\fill}
     \begin{minipage}[t]{8cm}
     \begin{center}\hspace{1cm}(b)\end{center}
     \epsfig{file=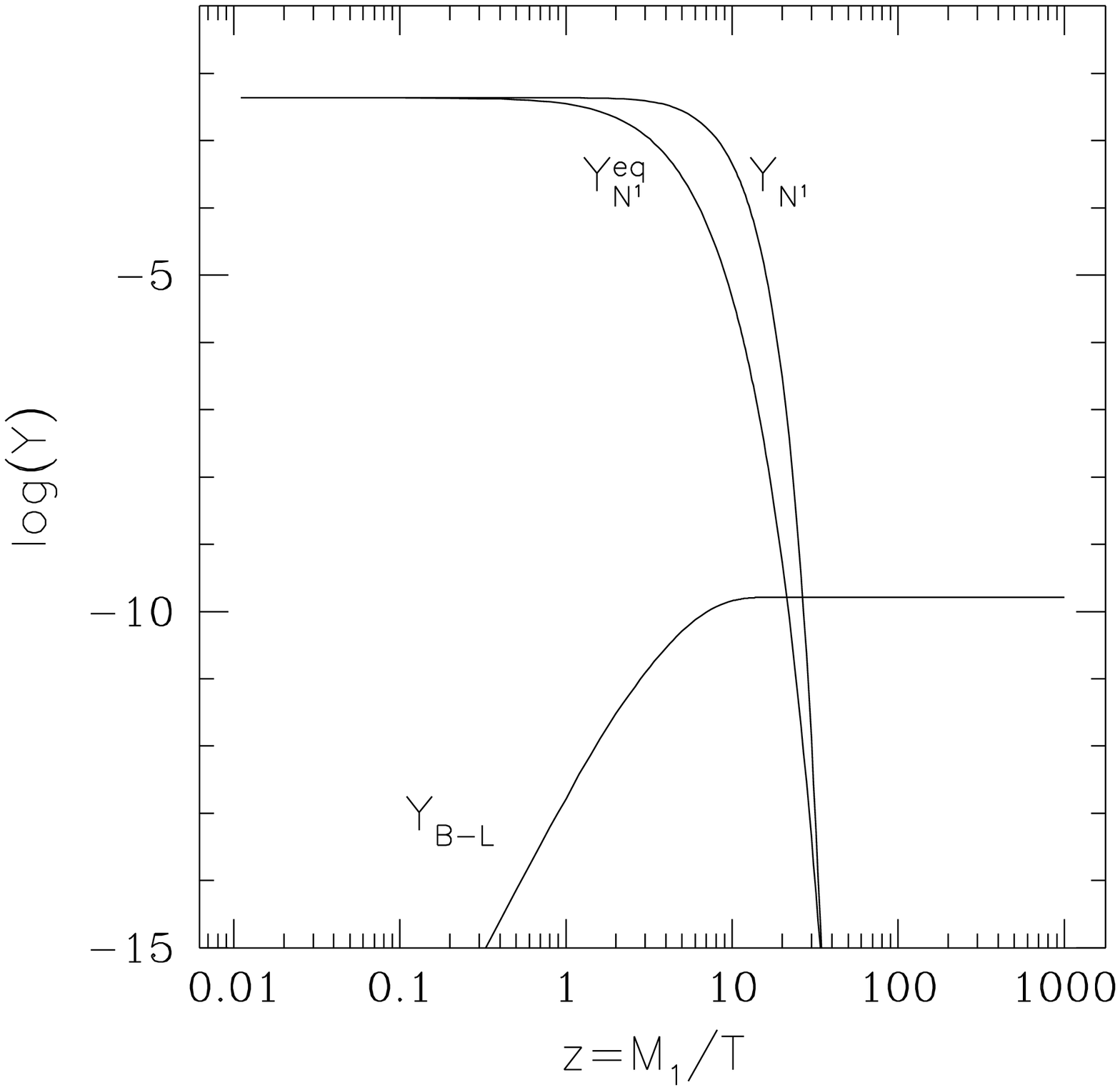,%
             bbllx=44pt,bblly=170pt,%
             bburx=560pt,bbury=690pt,width=8cm}
     \end{minipage}  
     \caption{\it Typical solutions of the Boltzmann equations.
       \label{sol1}}
    \end{figure}

    For simplicity we begin by neglecting the decays of the
    right-handed neutrinos $N^2$ and $N^3$. Using the formalism of
    section \ref{kinetic} and the reaction rates calculated in section
    \ref{model} we can immediately write down the Boltzmann equations.
    If we assume that all the standard model particles are in thermal
    equilibrium, the Boltzmann equations for the $N^1$ number and for
    the $(B-L)$ asymmetry are
    \beqa
    {\mbox{d}Y_{N^1}\over\mbox{d}z}&=&-{z\over sH\left(M_1\right)}
    \left\{\left({Y_{N^1}\over Y_{N^1}^{eq}}-1\right)\left[
    \g_{D1}+2\g_{\f,s}^1+4\g_{\f,t}^1\right]+\left[\left(
    {Y_{N^1}\over Y_{N^1}^{eq}}\right)^2-1\right]\g_{Z'}\right\}
    \label{50}\;,\\[1ex]
    {\mbox{d}Y_{B-L}\over\mbox{d}z}&=&-{z\over sH\left(M_1\right)}
    \left\{\left[{1\over2}{Y_{B-L}\over Y_l^{eq}}+
    \ve_1\left({Y_{N^1}\over Y_{N^1}^{eq}}-1\right)\right]\g_{D1}
    +\right.\NO\\[1ex]
    &&\qquad\left.+{Y_{B-L}\over Y_l^{eq}}\left[2\g_{N}+2\g_{N,t}+
    2\g_{\f,t}^1+{Y_{N^1}\over Y_{N^1}^{eq}}\g_{\f,s}^1\right]\right\}\;.
    \label{51}
    \eeqa
    These equations have to be solved numerically. We start at $T\gg
    M_1$ with the initial conditions
    \beq
     Y_{N^1}=Y_{N^1}^{eq}\qquad\mbox{and}\qquad Y_{B-L}=0\label{52}
    \eeq
    and follow the evolution of these quantities down to low
    temperatures. We want to emphasize that the results are
    independent of the initial conditions (\ref{52}), because the
    reaction rates are so high that the neutrinos are rapidly driven
    into equilibrium at high temperatures while any primordial $(B-L)$
    asymmetry is washed out. Typical results are shown in fig.~\ref{sol1},
    where we have used the following parameters
    \beqa
     &&M_1=10^{10}\,\mbox{GeV}\qquad\qquad m_{Z'}=10^{13}
     \,\mbox{GeV}\qquad\qquad \ve_1=-5\cdot10^{-8}\\[1ex]
     &&\mbox{and }\qquad 
     \tilde{m}_1:={(m_D^{\dg}m_D)_{11}\over M_1}=\left\{
     \begin{array}{l@{\qquad}l}
     10^{-6}\,\mbox{eV}&\mbox{for fig.~\ref{sol1}a}\\[1ex]  
     10^{-4}\,\mbox{eV}&\mbox{for fig.~\ref{sol1}b}\;.
     \end{array}\right.
    \eeqa
    \begin{figure}[t]
     \begin{minipage}[t]{8cm}
     \begin{center}\hspace{1cm}(a)\end{center}
     \epsfig{file=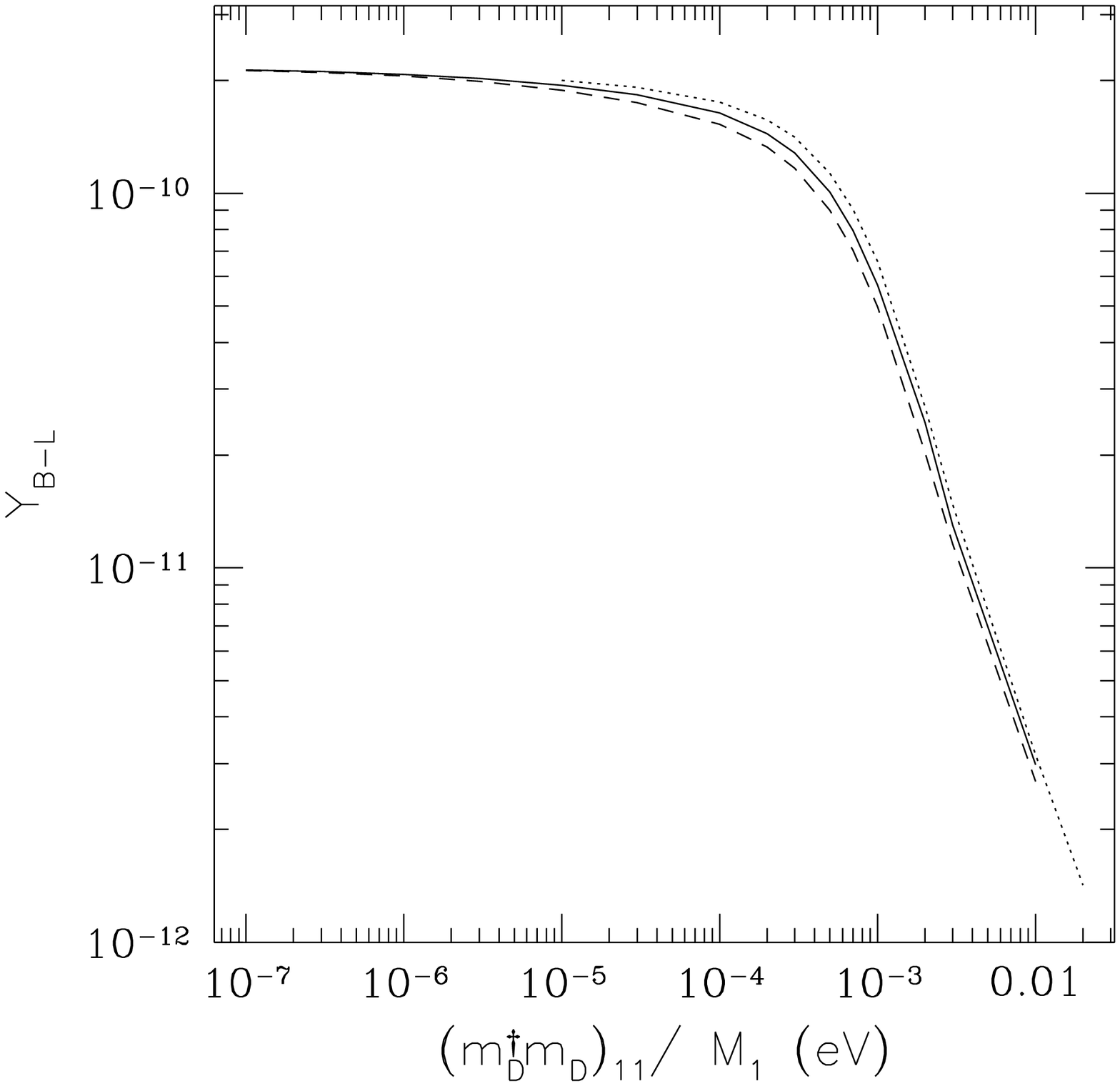,%
             bbllx=44pt,bblly=170pt,%
             bburx=570pt,bbury=700pt,width=8cm}
     \end{minipage}
     \hspace{\fill}
     \begin{minipage}[t]{8cm}
     \begin{center}\hspace{1cm}(b)\end{center}
     \epsfig{file=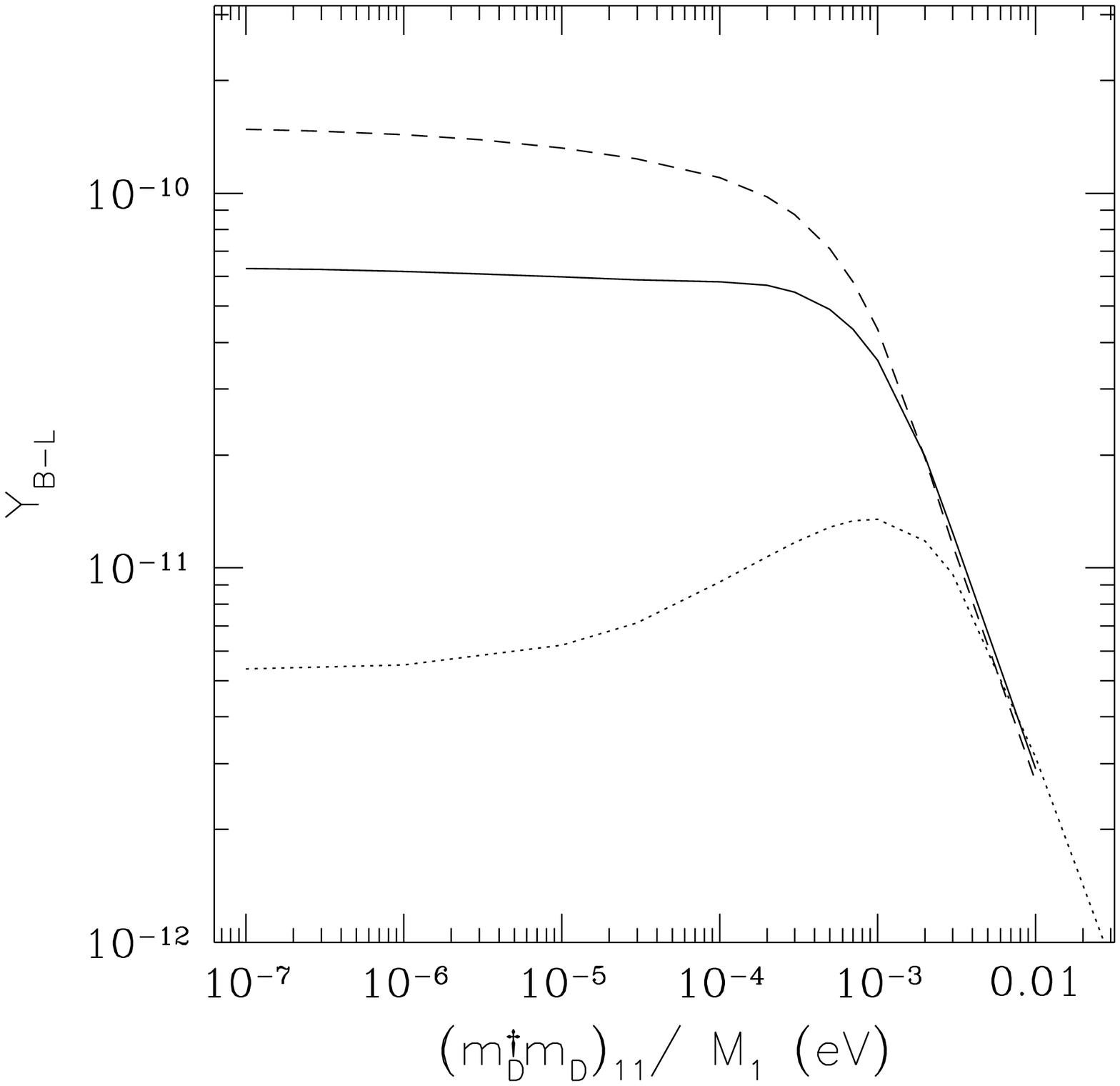,%
             bbllx=44pt,bblly=170pt,%
             bburx=570pt,bbury=700pt,width=8cm}
     \end{minipage} 
     \caption{\it The generated $(B-L)$ asymmetry for $m_{Z'}=10^3M_1$
       (a) and $m_{Z'}=10M_1$ (b) and for $M_1=10^8\,$GeV
       (dotted line), $M_1=10^{10}\,$GeV (solid line)
       and $M_1=10^{12}\,$GeV (dashed line). \label{sol2}} 
    \end{figure}
    One recognizes very clearly the departure from thermal equilibrium 
    and the generation of the asymmetry in the correct order of
    magnitude. If the neutrinos decay at very low temperatures $T\ll
    M_1$, the back reactions and the lepton number violating
    scatterings can be neglected, so that we expect a $(B-L)$
    asymmetry of the order \cite{kt1}
    \beq
     Y_{B-L}\approx-\ve_1 Y_{N^1}(T=M_1)=2.1\cdot10^{-10}\;.\label{55}
    \eeq
    This is exactly the result of fig.~\ref{sol1}a, while one has
    \beq
     Y_{B-L}=1.6\cdot10^{-10}\label{56}
    \eeq
    in fig.~\ref{sol1}b. This means that the inverse decays and the 
    lepton number violating scatterings already erase a part of the
    asymmetry. A further increase in $(m_D^{\dg}m_D)_{11}$ would
    intensify the dissipative processes, so that the final asymmetry
    would be even lower.

    One gets almost the same results if one varies the parameters and
    keeps the ratios $y=(m_{Z'}/M_1)^2$ and $\tilde{m}_1$ and the $CP$
    asymmetry $\ve_1$ constant. In fig.~\ref{sol2}a we have plotted the
    generated asymmetry as a function of $\tilde{m}_1$ for $y=10^6$
    and $\ve_1=-5\cdot10^{-8}$ for the neutrino masses
    $M_1=10^{12}\,$GeV, $10^{10}\,$GeV and $10^8\,$GeV.

    First of all one notices that the condition (\ref{45}) delimitates
    quite accurately the range of $\tilde{m}_1$ in which the
    generation of the $(B-L)$ asymmetry is possible with a reasonable
    choice for the $CP$ asymmetry.

    In the second place it is remarkable that the asymmetry depends
    almost entirely on $\tilde{m}_1$ and not on $(m_D^{\dg}m_D)_{11}$
    and $M_1$ separately. This becomes clear if one looks at the
    right-hand side of the Boltzmann equation (\ref{51}). Indeed the
    decay term depends only on $\tilde{m}_1$,
    \beq
    {z\over sH\left(M_1\right)}\,\g_{D1}\propto\tilde{m}_1\;.
    \eeq
    The same proportionality holds for $\g_{\f,s}$ and
    $\g_{\f,t}$. On the other hand one finds
    \beq
    {z\over sH\left(M_1\right)}\,\g_{N}\propto\tilde{m}_1^2\,M_1
    \qquad\mbox{and}\qquad
    {z\over sH\left(M_1\right)}\,\g_{N,t}\propto\tilde{m}_1^2\,M_1 
     \eeq
    for the lepton number violating scatterings. Since these terms can
    erase a lepton asymmetry one would expect that the asymmetry is
    falling with $M_1$. This is exactly what one can observe in 
    fig.~\ref{sol2}a.

    Up to now we have always chosen a large $Z'$ mass,
    $m_{Z'}=10^3M_1$, to ensure that pair annihilation processes do
    not influence the evolution of $Y_{N^1}$ at temperatures $T<M_1$.
    As a counter-example we have repeated our calculations with a
    lower $Z'$ mass, $m_{Z'}=10M_1$. The results are summarized in
    fig.~\ref{sol2}b. In contrast to fig.~\ref{sol2}a one recognizes a
    strong dependence of the asymmetry on $M_1$. This comes about
    because of the pair annihilation term in the Boltzmann equation
    (\ref{50}), which is inversely proportional to $M_1$,
    \beq
    {z\over sH\left(M_1\right)}\,\g_{Z'}\propto {1\over M_1}\;.
    \eeq
    Therefore, the pair annihilation processes should be more effective
    at low neutrino masses $M_1$. This explains the dependence of the
    curves in fig.~\ref{sol2}b on $M_1$. 

    Fig.~\ref{sol2}b may further be used to confirm the lower bound
    (\ref{46}) on the $Z'$ mass. One can generate the required
    asymmetry for $M_1=10^{12}$ and $10^{10}\,$GeV but not for
    $M_1=10^8\,$GeV, because the condition (\ref{46}) is only
    fulfilled for the first two values of $M_1$.

    The curve for $M_1=10^8\,$GeV in fig.~\ref{sol2}b has another
    interesting property. Up to now the generated asymmetry was always
    monotonically falling with $\tilde{m}_1$, while $Y_{B-L}$ has now
    a maximum for $\tilde{m}_1\approx10^{-3}\,$eV. The reason is, that
    for this value of $\tilde{m}_1$ the neutrinos decay before they
    can be depleted by pair annihilation processes, while the reaction
    rate for inverse decays is still too small to erase the asymmetry.
    The evolution of the asymmetry is hampered by pair annihilation
    processes at lower values of $\tilde{m}_1$ and by dissipating
    processes like inverse decays at higher values of $\tilde{m}_1$.
\subsection{Decays of two heavy neutrinos}
    Now we can refine our results by adding a second family of
    right-handed neutrinos. The Boltzmann equations are
    straightforward generalizations of the equations (\ref{50}) and
    (\ref{51})
    \beqa
    {\mbox{d}Y_{B-L}\over\mbox{d}z}&=&-{z\over sH\left(M_1\right)}
    \left\{\sum\limits_{j=1}^2\left[{1\over2}{Y_{B-L}\over Y_l^{eq}}+
    \ve_j\left({Y_{N^j}\over Y_{N^j}^{eq}}-1\right)\right]\g_{Dj}
    \right.\NO\\[1ex]
    &&\left.+{Y_{B-L}\over Y_l^{eq}}\left[2\g_{N}+2\g_{N,t}\right]
    +{Y_{B-L}\over Y_l^{eq}}\sum\limits_{j=1}^2\left[2\g_{\f,t}^j
    +{Y_{N^j}\over Y_{N^j}^{eq}}\g_{\f,s}^j\right]\right\}\;,\\[1ex]
    {\mbox{d}Y_{N^1}\over\mbox{d}z}&=&-{z\over sH\left(M_1\right)}
    \left\{\left({Y_{N^1}\over Y_{N^1}^{eq}}-1\right)\left[
    \g_{D1}+2\g_{\f,s}^1+4\g_{\f,t}^1\right]+\right.\NO\\[1ex]
    &&+\left.\left[\left({Y_{N^1}\over Y_{N^1}^{eq}}\right)^2-
    1\right]\g_{Z'}+\left[\left({Y_{N^1}\over Y_{N^1}^{eq}}\right)^2
    -\left({Y_{N^2}\over Y_{N^2}^{eq}}\right)^2\right]\g_{N^1N^2}
    \right\}\;,\\[1ex]
    {\mbox{d}Y_{N^2}\over\mbox{d}z}&=&-{z\over sH\left(M_1\right)}
    \left\{\left({Y_{N^2}\over Y_{N^2}^{eq}}-1\right)\left[
    \g_{D2}+2\g_{\f,s}^2+4\g_{\f,t}^2\right]+\right.\NO\\[1ex]
    &&+\left.\left[\left({Y_{N^2}\over Y_{N^2}^{eq}}\right)^2-
    1\right]\g_{Z'}+\left[\left({Y_{N^2}\over Y_{N^2}^{eq}}\right)^2
    -\left({Y_{N^1}\over Y_{N^1}^{eq}}\right)^2\right]\g_{N^1N^2}
    \right\}\;,
    \eeqa
    where $\g_{N^1N^2}$ is the reaction density for the
    $N^1$-$N^2$-transitions described by the reduced cross section
    (\ref{43}).

    The constraints which were derived in section \ref{constraints}
    remain valid, but the parameters for the two heavy neutrinos are
    not always independent of each other. For the $CP$ asymmetries one
    has for example:
    \beq
     \ve_1\propto\mbox{Im}\left[\left(m_D^{\dg}m_D\right)_{12}^2
     \right]\qquad\mbox{and}\quad
     \ve_2\propto\mbox{Im}\left[\left(m_D^{\dg}m_D\right)_{21}^2
     \right]\;.
    \eeq
    Since 
    \beq
     \left(m_D^{\dg}m_D\right)_{12}=\left(m_D^{\dg}m_D\right)_{21}^*\;,
    \eeq
    the two $CP$ asymmetries have a different sign. Therefore, the
    asymmetries generated in the decays of $N^1$ and $N^2$ will also
    have a different sign, so that they can cancel each other if the
    neutrinos are mass degenerate. But this should not be a problem if
    the neutrinos have hierarchical masses like the charged fermions,
    because the asymmetry generated by the heavier neutrinos is more
    affected by the dissipative processes and because the mass
    dependence of $\ve_1$ and $\ve_2$ ensures that $\abs{\ve_2}$ is
    smaller than $\abs{\ve_1}$ if $M_2>M_1$.
    Analogous phenomena have already been observed in other models 
    \cite{hkrw}.

    \begin{figure}[t]
     \begin{minipage}[t]{8cm}
     \begin{center}\hspace{1cm}(a)\end{center}
     \epsfig{file=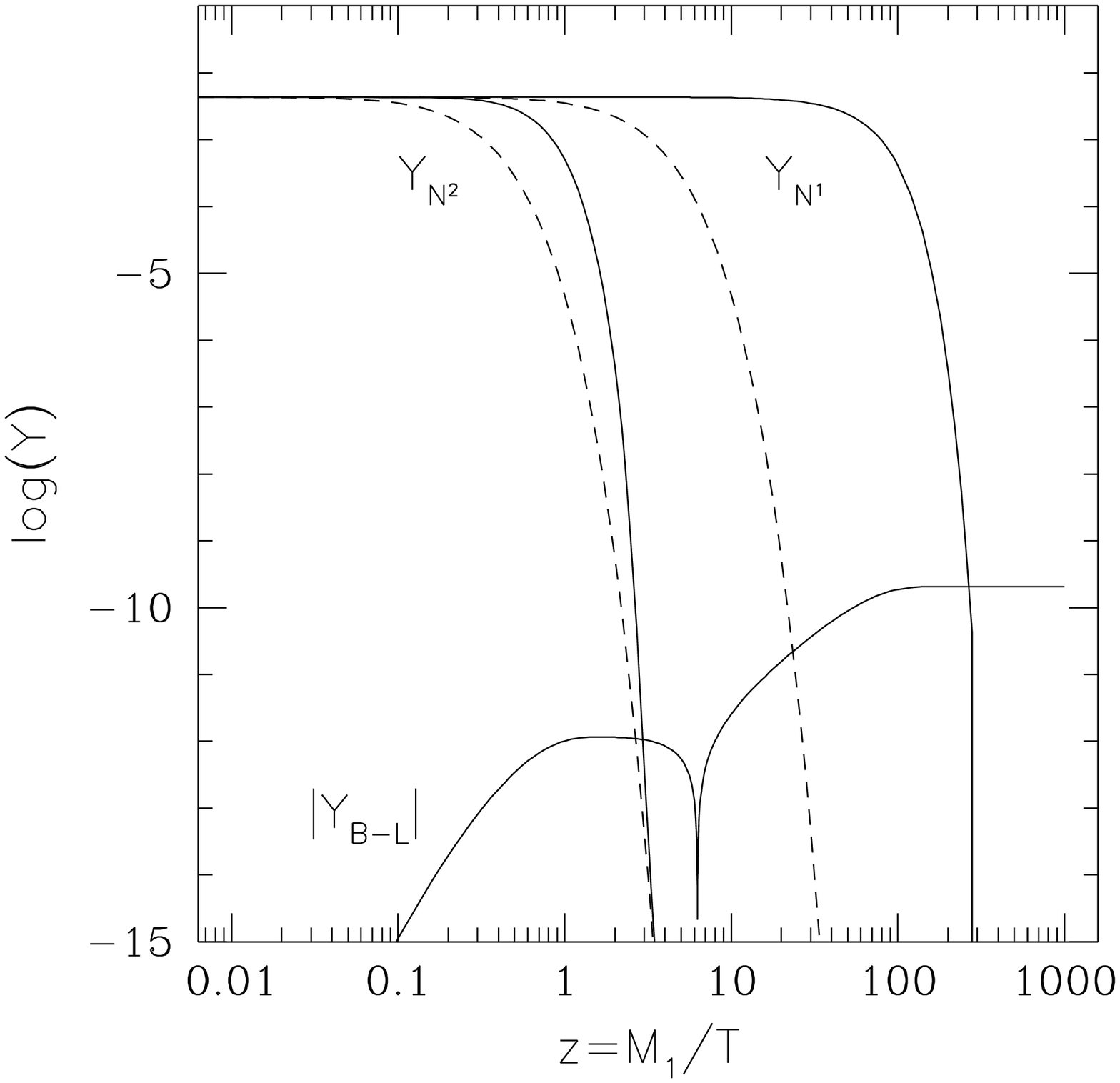,%
             bbllx=44pt,bblly=170pt,%
             bburx=560pt,bbury=690pt,width=8cm}
     \end{minipage}
     \hspace{\fill}
     \begin{minipage}[t]{8cm}
     \begin{center}\hspace{1cm}(b)\end{center}
     \epsfig{file=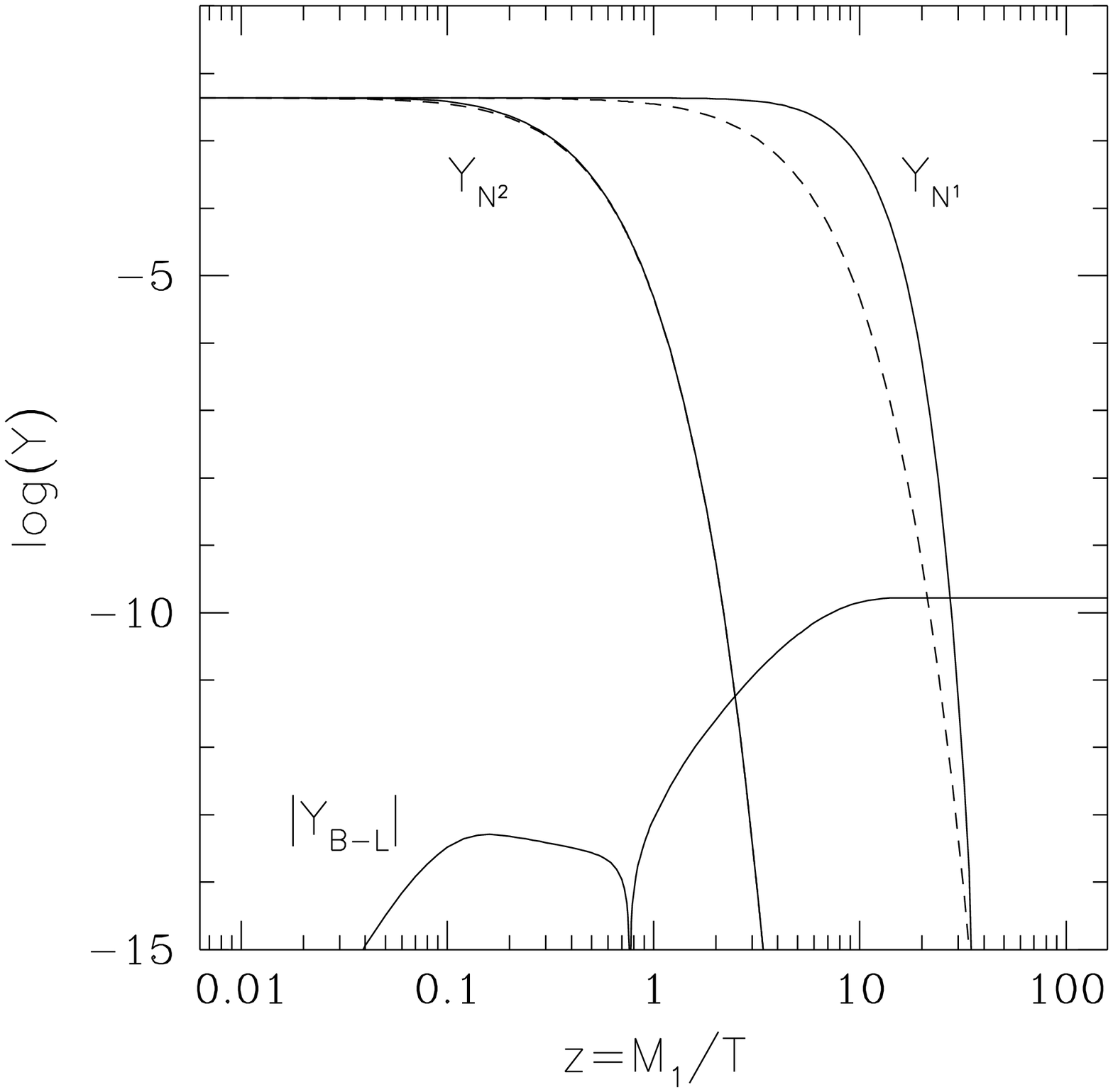,%
             bbllx=44pt,bblly=170pt,%
             bburx=560pt,bbury=690pt,width=8cm}
     \end{minipage}  
     \caption{\it Typical solutions of the Boltzmann equations with
       two neutrino families. The equilibrium distributions are
       represented by dashed lines. \label{sol3}}
    \end{figure}
    Typical results of the numerical integration of the Boltzmann
    equations are shown in fig.~\ref{sol3}. In both cases we have chosen
    \beq
     M_1=10^{10}\,\mbox{GeV}\;,\qquad M_2=10^{11}\,\mbox{GeV}\;,
     \qquad M_{Z'}=10^{13}\,\mbox{GeV}\quad\mbox{and}\quad
     \mbox{Re}\left[\left(m_D^{\dg}m_D\right)_{12}^2\right]=0.
    \eeq
    The results are almost independent of the choice 
    $\mbox{Re}\left[\left(m_D^{\dg}m_D\right)_{12}^2\right]=0$,
    because this parameter only appears in interference terms between
    the different diagrams contributing to the lepton Higgs
    scatterings which have only a small influence on the generated
    asymmetry. 

    Furthermore we have for fig.~\ref{sol3}a
    \beqa
     &&\tilde{m}_1={\tilde{m}_2\over100}=10^{-6}\,\mbox{eV}
     \qquad\mbox{and}\qquad
     {\mbox{Im}\left[\left(m_D^{\dg}m_D\right)_{12}^2\right]\over
     M_1^2}=7.7\cdot10^{-8}\,(\mbox{eV})^2\NO\\[1ex]
     &&\Leftrightarrow\qquad\ve_1=-5\cdot10^{-8}\qquad\mbox{and}
     \qquad\ve_2=3.7\cdot10^{-10}\;,
    \eeqa
    and for fig.~\ref{sol3}b we have
    \beqa
     &&\tilde{m}_1={\tilde{m}_2\over100}=10^{-4}\,\mbox{eV}
     \qquad\mbox{and}\qquad
     {\mbox{Im}\left[\left(m_D^{\dg}m_D\right)_{12}^2\right]\over
     M_1^2}=7.7\cdot10^{-6}\,(\mbox{eV})^2\NO\\[1ex]
     &&\Leftrightarrow\qquad\ve_1=-5\cdot10^{-8}\qquad\mbox{and}
     \qquad\ve_2=3.7\cdot10^{-10}\;.
    \eeqa
    Therefore, we have chosen exactly the same parameters for $N^1$ as
    in the one-family calculations of fig.~\ref{sol1}, so that we can
    compare the results. First of all one sees that the asymmetry
    changes sign at $z\approx6$ and $z\approx0.8$ as we had predicted
    it at the beginning of this section. But the influence of the
    second neutrino family on the final asymmetry is only very small
    and we recover the results (\ref{55}) and (\ref{56}). Consequently
    one can always neglect the heavier neutrino families if one has a
    pronounced mass hierarchy.
\subsection{Physical mass matrices}
    Up to now we have only verified that the parameters of the theory can
    be chosen in such a way, that they allow the generation of the 
    requested asymmetry. However, in a physical theory the parameters
    have to comply with some other conditions. First of all the mass
    scale of the right-handed neutrinos and the $Z'$ boson have to be
    related to the intermediate breaking scale $v'$. Next the
    predicted light neutrino masses have to be consistent with the 
    experimental bounds. We will focus on the following conditions:
    \begin{enumerate}
       \item The solar neutrino deficit can be explained by the
         Mikheyev-Smirnov-Wolfenstein (MSW) effect, if the $\n_e$ and
         $\n_{\m}$ masses and their mixing angle fulfill the following 
         conditions (cf.~\cite{fy1})
         \beqa
         &&\d m^2 := m_{\n_{\m}}^2-m_{\n_e}^2=(0.3-1)\cdot10^{-5}
         \,(\mbox{eV})^2\label{68}\;,\\[1ex]
         &&\sin^22\q_{e\m}=0.003-0.012\label{69}\;;
         \eeqa
       \item a $\n_{\t}$ mass of a few eV is needed in
         the cold-plus-hot dark matter models.
     \end{enumerate}
     A realistic pattern of masses and mixings of the known fermions
     can be obtained based on a gauged abelian family symmetry broken
     below the unification scale \cite{ross}. This model predicts the
     following mass matrices for the up and down quarks (suppressing
     unknown phases and factors which are assumed to be of order 1)
     \beq
     M_u\approx \left(
     \begin{array}{ccc}
     \e^{\mid 2+6a \mid } & \e^{\mid 3a \mid } & \e^{\mid 1+3a\mid }\\
     \e^{\mid 3a \mid } & \e^2 & \e \\
     \e^{\mid 1+3a \mid } & \e & 1
     \end{array}\right)m_t\qquad\mbox{and}\qquad
     M_d\approx \left (
     \begin{array}{ccc}
     \bar{\e}^{\mid 2+6a \mid } & \bar{\e}^{\mid 3a \mid } & 
     \bar{\e}^{\mid 1+3a \mid } \\
     \bar{\e}^{\mid 3a \mid } & \bar{\e}^2 & \bar{\e} \\
     \bar{\e}^{\mid 1+3a \mid } & \bar{\e} & 1
     \end{array}\right)m_b\;,
     \eeq
     with $a=\a_3/(\a_2-\a_3)$, where $\a_i$ is the family charge of
     the $i$-th generation quarks. One obtains a good agreement with the
     measured values of the quark masses and mixings for
     \beq
     a=1\qquad\mbox{and}\qquad\sqrt{\e}=\bar{\e}=0.23\;.
     \eeq
     Similarly the leptons of the $i$-th generation have the family
     charge $a_i$, and therefore the mass matrix of the charged leptons
     reads
     \beqa
     M_l&\approx& \left (
     \begin{array}{ccc}
     \bar{\e}^{\mid 8-2b \mid } & \bar{\e}^3&
     \bar{\e}^{\mid 4-b \mid } \\
     \bar{\e}^3 & \bar{\e}^{ \mid 2(1-b) \mid }&
     \bar{\e}^{ \mid 1-b \mid} \\
     \bar{\e}^{\mid 4-b \mid } & \bar{\e}^{\mid 1-b \mid} & 1
     \end{array}\right)m_{\t}
     \qquad\mbox{for integer }b={\a_2-a_2\over\a_2-\a_3}\;.
     \eeqa
     The known masses of the charged leptons can be explained for
     $b=0$. $M_l$ is diagonalized by an orthogonal matrix $R_l$. If
     one neglects higher order terms in $\bar{\e}$, $R_l$ is given by
     \beq
     R_l\approx\left(
     \begin{array}{ccc}
     1 &  \d_{e\m} & \co\left(\bar{\e}^4\right) \\
     -\d_{e\m} & 1 & \co\left(\bar{\e}\right) \\
     -\co\left(\bar{\e}^4\right) & -\co\left(\bar{\e}\right) & 1
     \end{array}\right)\qquad\mbox{with }\d_{e\m}=\sqrt{m_e\over m_{\m}}
     =\co(\bar{\e}^2)\;.
     \eeq
     This model also predicts Dirac masses $m_D$ for the neutrinos and
     Majorana masses $M$ for the right-handed neutrinos
     \cite{dreiner},
     \beq
     m_D= \left (
     \begin{array}{ccc}
     \bar{\e}^{16} & \bar{\e}^6 & \bar{\e}^8 \\
     \bar{\e}^6 & \bar{\e}^4 & \s\bar{\e}^2 \\
     \bar{\e}^8 & \s\bar{\e}^2 & \r
     \end{array}\right)m_{\nu_3}
     \qquad\mbox{and}\qquad
     M= \left (
     \begin{array}{ccc}
     \bar{\e}^8 & \t\bar{\e}^3 & \bar{\e}^4 \\
     \t\bar{\e}^3 & \z\bar{\e}^2 & \bar{\e} \\
     \bar{\e}^4 & \bar{\e} & 1
     \end{array}\right)M_3\;,
     \eeq
     where we have introduced four factors $\r$, $\s$, $\t$ and $\z$ of
     order 1, which will be used to fix the physical parameters. As we
     are working with mass eigenstates $N^i$ we have to transform
     $m_D$ and $M$ into a basis in which $M$ is diagonal. Up to higher
     order terms in $\bar{\e}$ the matrix $R_M$ which diagonalizes $M$
     is
     \beq
     R_M=\left(
     \begin{array}{ccc}
     1 & \bar{\e}^2  & \bar{\e}^4 \\
     -\bar{\e}^2 & 1 & \bar{\e} \\
     -\bar{\e}^4 & -\bar{\e} & 1
     \end{array}\right)\;,
     \eeq
     which yields the following masses for the heavy neutrinos
     \beq
     M' := R_M^TMR_M\approx\left(
     \begin{array}{ccc}
     -2\t\bar{\e}^5 & 0 & 0 \\
     0 & (\z-1)\bar{\e}^2 & 0 \\
     0 & 0 & 1
     \end{array}\right)M_3\;.
     \eeq
     Therefore the mass ratios of the right-handed neutrinos are
     \beq
     a_2=\left({M_2\over M_1}\right)^2\approx{(\z-1)^2\over4\t^2\bar{\e}^6}
     \qquad\mbox{and}\qquad
     a_3=\left({M_3\over M_1}\right)^2\approx{1\over4\t^2\bar{\e}^{10}}\;.
     \eeq
     The Dirac mass matrix in the new basis is
     \beq
     m_D':=m_D\,R_M\approx\left(
     \begin{array}{ccc}
     -\bar{\e}^8 & \bar{\e}^6 & \bar{\e}^7 \\
     -(1+\s)\bar{\e}^6 & -\s\bar{\e}^3 & \s\bar{\e}^2 \\
     -(\r+\s)\bar{\e}^4 & -(\r-\s\bar{\e})\bar{\e} & \r
     \end{array}\right)m_{\nu_3}\;.
     \eeq
     \begin{figure}[t]
     \begin{center}
     \epsfig{file=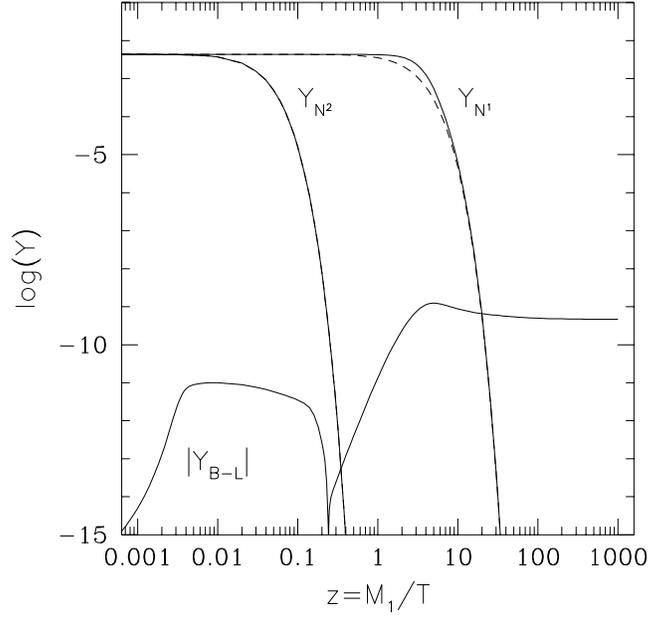,%
             bbllx=44pt,bblly=210pt,%
             bburx=560pt,bbury=690pt,width=8cm}
     \end{center}
     \caption{\it Solutions of the Boltzmann equations with physical
       mass matrices. The equilibrium distributions are
       represented by dashed lines.  \label{sol4}}
    \end{figure}
     The mass parameters which appear in the decay widths are thus
     given by
     \beq
     \tilde{m}_1={\left({m'}_D^{\dg}m'_D\right)_{11}\over M_1}\approx
     {(\r+\s)^2\bar{\e}^2\over2\t}{m_{\nu_3}^2\over M_3}
     \qquad\mbox{and}\qquad
     \tilde{m}_2={\left({m'}_D^{\dg}m'_D\right)_{22}\over M_2}\approx
     {(\r-\s\bar{\e})^2\over\abs{\z-1}}{m_{\nu_3}^2\over M_3}\;.\label{78}
     \eeq
     Up to now we have neglected all the phase factors that we need
     for $CP$ violation. In a Majorana mass matrix the number of
     physical phases for the $n$ generation case is $n(n-1)/2$
     (cf.\ \cite{fy1}). They can be chosen in such a way, that the squared
     non diagonal elements of ${m'}_D^{\dg}m'_D$ are purely
     imaginary. Then the $CP$ asymmetries are
     \beq
     \ve_1\approx \t\left[\r^2+{(\r-\s\bar{\e})^2\over\abs{\z-1}}\right]
     \left({m_{\nu_3}\over3.6\cdot10^4\,\mbox{GeV}}\right)^2
     \qquad\mbox{and}\qquad\ve_2\approx \r^2\abs{\z-1}
     \left({m_{\nu_3}\over5.6\cdot10^3\,\mbox{GeV}}\right)^2\;.\label{79}
     \eeq
     Now we have to specify the parameters $\r$, $\s$, $\t$ and
     $\z$. A good choice is
     \beq
     \r=0.9\;,\qquad \s=-1\;,\qquad \t=-{1\over4}\quad\mbox{and}\quad\z=2\;.
     \eeq
     The mass scales are approximately fixed by their breaking scale,
     \beq
     m_{\n_3}=50\,\mbox{GeV}\;,\qquad M_3=5\cdot10^{11}\,\mbox{GeV}
     \quad\mbox{and}\quad M_{Z'}=10^{12}\,\mbox{GeV}\;.
     \eeq
     Then the Majorana neutrinos $N^1$ and $N^2$ have the masses
     \beq
     M_1=3.1\cdot10^8\,\mbox{GeV}\qquad\mbox{and}\qquad
     M_2=2.6\cdot10^{10}\,\mbox{GeV}\;.
     \eeq
     For the coupling parameters (\ref{78}) and the $CP$ asymmetries
     (\ref{79}) one finds
     \beqa
     &&\tilde{m}_1=
     8\cdot10^{-4}\,\mbox{eV}\qquad\mbox{and}\qquad
     \tilde{m}_2=
     6.4\,\mbox{eV}\label{3_76}\;,\\[1ex]
     &&\ve_1=-1.3\cdot10^{-6}\qquad\mbox{and}\qquad
     \ve_2=4.6\cdot10^{-5}\;.
     \eeqa
     The corresponding solutions of the Boltzmann equations are shown
     in fig.~\ref{sol4}. The generated asymmetry,
     \beq
     Y_{B-L}=5\cdot10^{-10}\;,
     \eeq
     is even bigger than requested, but it is always possible to
     reduce $Y_{B-L}$ by choosing the phase factors in an appropriate
     way. One finds similar results for other values of the parameters
     $\r$, $\s$, $\t$ and $\z$.

     Finally we have to check if these matrices can predict the
     desired light neutrino masses and mixings. The Majorana masses of
     the light neutrinos are given by the seesaw formula \cite{seesaw},
     \beq
     m_{\n}=m_D{1\over M}m_D^T\approx\left(
     \begin{array}{ccc}
     -2\bar{\e}^{15} & (\s-1)\bar{\e}^9 & -\r\bar{\e}^7 \\
     (\s-1)\,\bar{\e}^9 & \s^2\t\,\bar{\e}^4 & \r\s\t\,\bar{\e}^2 \\
     -\r\,\bar{\e}^7 & \r\s\t\,\bar{\e}^2 & \r^2\t
     \end{array}\right)\,{1\over\t}\,{m_{\nu_3}^2\over M_3}\;.
     \eeq
     The eigenvalues of this matrix are
     \beq
     m_{\n_e}=8.9\cdot10^{-7}\,\mbox{eV}\;,\qquad
     m_{\n_{\m}}=1.9\cdot10^{-3}\,\mbox{eV}\quad\mbox{and}\quad
     m_{\n_{\t}}=2.2\,\mbox{eV}\;.
     \eeq
     These values fulfill the MSW condition (\ref{68}) and the
     $\n_{\t}$ mass is in the correct range for the cold-plus-hot dark
     matter models. $m_{\n}$ is approximately diagonalized by the
     following matrix,
     \beq
     R_{\n}\approx\left(
     \begin{array}{ccc}
     1 & 2\bar{\e}^3 & \bar{\e}^5 \\
     -2\bar{\e}^3 & 1 &\displaystyle  (\s/\r)\bar{\e}^2 \\
     -\bar{\e}^5 &\displaystyle  -(\s/\r)\bar{\e}^2 & 1
     \end{array}\right)\;.
     \eeq
     The leptonic analogon of the Cabbibo-Kobayashi-Maskawa
     (CKM) matrix is
     \beq
     V_l=\left(R_{\n}\right)^{-1}R_l=\left(
     \begin{array}{ccc}
     1 &  \d_{e\m}-2\bar{\e}^3 & -\bar{\e}^4 \\
     -\d_{e\m}+2\bar{\e}^3 & 1 & \bar{\e} \\
     \bar{\e}^4 & -\bar{\e} & 1
     \end{array}\right)\;.
     \eeq
     Therefore the $\n_e$-$\n_{\m}$ mixing angle complies with the MSW
     condition (\ref{69})
     \beq
     \sin^22\q_{e\m}=0.009\;.
     \eeq
\section{Conclusions}
    We have seen that the cosmic baryon asymmetry can be explained by
    the lepton number violating decays of heavy Majorana neutrinos
    combined with the anomalous electroweak $(B+L)$ violation. The
    lepton asymmetry generated in our model is independent of the
    initial conditions on the heavy neutrino density, which was not
    the case in a previous analysis of this mechanism \cite{luty}.
    This is a consequence of the new gauge interaction which we
    have introduced, and which is related to the spontaneous breaking
    of lepton number.

    We have also considered the decay of more than one heavy neutrino
    and have seen that the generated asymmetry is determined by the
    properties of the lightest heavy neutrino, if the right-handed
    neutrinos have a pronounced mass hierarchy.

    By performing explicit calculations we could show that the
    generation of a lepton asymmetry is possible at every temperature
    between the intermediate breaking scale and the electroweak
    breaking scale. Furthermore we have checked that neutrino mass
    matrices may explain the generation of the asymmetry and low
    energy neutrino phenomenology at the same time.

    In supersymmetric models of inflation the reheating temperature
    has to be lower than $10^5$ to $10^8\,$GeV to solve the gravitino
    problem \cite{sarkar,camp}. Therefore in these models
    baryogenesis has to take place at relatively low
    temperatures. Since this is possible in our model, 
    a supersymmetric generalization should be viable.
\vspace{1cm}\mbox{ }\\    
\noindent
\setlength{\parskip}{1ex}
{\bf\Large Acknowledgments}

\mbox{ }\\\noindent
I would like to thank W.~Buchm\"uller, who suggested this
investigation, for continuous support and encouragement.

\end{document}